# On the metric hypercomplex group alternative-elastic algebras for n mod 8 = 0.

K.V. Andreev

In this article the fine-tuning of Algorithm 9.1 [2], [3] is considered. In this connection, answers to the following questions are given.

1. How to construct the metric hypercomplex Cayley-Dickson algebra by means of Algorithm 9.1 for $n = 2^k$?
2. How to construct the metric hypercomplex orthogonal group alternative-elastic algebra by means of Algorithm 9.1 for $n \textit{ mod } 8 \ = 0$?
3. How to decompose the metric hypercomplex orthogonal homogenous group alternative-elastic algebra on an algebraic basis?
4. What is the generator of the metric hypercomplex orthogonal homogenous group alternative-elastic algebra and how to construct it?
5. How technically to realize Algorithm 9.1 and to construct the canonical sedenion algebra for n=16?
6. How to associate the metric hypercomplex orthogonal homogenous group alternative-elastic algebra to the spinor quadratic form?

Let *product* of elements of *hypercomplex group alternative-elastic algebra* $\mathbb{A}$ [1],[5], [2], [3] over the field $\mathbb{R}$ and the vector space $\mathbb{R}^n$ (n mod 8=0) be defined as

**Axiom 1.** *For any elements $a$ and $b$, their product $c$ is uniquely defined:*

$$c = ab.$$

**Axiom 2.** *There exists the unique identity element $e$. For any element $a$:*

$$ae = ea = a.$$

**Axiom 3.** *For any element $a \neq 0$, the inverse element $a^{-1}$ is uniquely determined:*

$$aa^{-1} = a^{-1}a = e.$$

**Axiom 4.** *(The weakly alternative identity.) For any elements $a$ and $b$:*

$$(aa)b - a(ab) = b(aa) - (ba)a.$$

**Axiom 5.** *(The flexible identity.) For any elements $a$ and $b$:*

$$a(ba) = (ab)a.$$

**Axiom 6.** *(The distributive identity.) For any elements $a, b$ and $c$:*

$$a(b + c) = ab + ac, \ \ (b + c)a = ba + ca.$$

Construct one of such the algebras. Suppose that the vector space $\mathbb{R}^n$ is equipped with the metric $(i, j, ... = \overline{1, n})$

$$< a, b > e := \tfrac{1}{2}(a\bar{b} + b\bar{a}), \quad a = a_0 e + \sum_{r=1}^{n-1} a_r e_r, \quad \bar{a} = a_0 e - \sum_{r=1}^{n-1} a_r e_r. \tag{1}$$

Let's consider that the metric $g_{ij}$ is Euclidean metric: $< a, b >= \delta_{ij} a^i b^j$ in the special orthogonal basis.

$$\begin{gathered} a^{-1} = \tfrac{\bar{a}}{<a,a>}, \quad \forall a \neq 0, \\ a + \bar{a} = 2a_0 e = 2 < a, e > e, \\ a(a + \bar{a}) = 2 < a, e > a, \\ a^2 = - < a, a > e + 2 < a, e > a, \\ (a + b)^2 = a^2 + b^2 + ab + ba = -(< a, a > + < b, b > +2 < a, b >)e+ \\ +2(< a, e > a+ < a, e > b+ < b, e > a+ < b, e > b)), \\ (a, b) := \tfrac{1}{2}(ab + ba) = - < a, b > e+ < a, e > b+ < b, e > a. \end{gathered} \tag{2}$$

**Definition 1.** *Let the identities*

$$(a, b) = - < a, b > e+ < a, e > b+ < b, e > a, \quad \eta_{(ij)}{}^k = 2(\frac{1}{\sqrt{2}}\eta_{(i})\delta_{j)}{}^k - g_{ij}(\frac{1}{\sqrt{2}}\eta^k), \tag{3a}$$

$$\frac{1}{2} < ab - ba, a >= 0, \quad \eta_{(i|j|k)} - \eta_{j(ik)} = 0, \tag{3b}$$

$$\frac{1}{2} < ab - ba, e >= 0, \quad \eta_{[ij]}{}^k(\frac{1}{\sqrt{2}}\eta_k) = 0. \tag{3b$'$}$$

*be executed by definition for all a and b, where $\eta_{ij}{}^k$ are the structural constants of the unital algebra with the algebra identity $e = \frac{1}{\sqrt{2}}\eta^k$ and the distributive identity. We call the algebra* ***orthogonal hypercomplex algebra*** $\mathbb{A}$.

**Theorem 1.** *The orthogonal hypercomplex algebra will be the metric hypercomplex group alternative-elastic algebra.*

*Proof.* The equations (3) are a consequence of the Clifford equation for n mod 8=0 [3, p. 37, eq. (8.1) (eng), p. 166(43), eq. (8.1) (rus)] $(A, B, ... = \overline{1, N}, \quad N = 2^{\frac{n}{2}-1})$

$$\eta_i{}^{AB}\eta_{jCB} + \eta_j{}^{AB}\eta_{iCB} = g_{ij}\delta_C{}^B. \tag{4}$$

At the same time, $\eta_i{}^{AB} = \sum_{I=0}^{n-1} \frac{1}{2}(\eta_I)_i(\varepsilon_I)^{AB}$. Among the tensors $(\varepsilon_I)^{AB}$, there is only the one symmetric tensor ($(\varepsilon_0)^{AB} =: \varepsilon^{AB}$ - the metric spin-tensor), the remaining tensors are antisymmetric tensors. Then the structure constants of the orthogonal hypercomplex algebra $\mathbb{A}$ have the form (Theorem 3)

$$\eta_{ij}{}^k := \sqrt{2}\eta_i{}^{AB}\eta_{jCA}\eta^k{}_{DB}\theta^{CD}, \quad \bar{X}^{A'} = S_A{}^{A'}X^A, \tag{5}$$

where $\theta^{CD}(\theta^{CD}\varepsilon_{CD} := 2)$ is an arbitrary symmetric tensor, and $S_A{}^{A'}$ is generated by the real inclusion $H_i{}^\Lambda : \mathbb{R}^n \longmapsto \mathbb{C}^n$ $(\Lambda, \Psi, ... = \overline{1, n})$ [3, pp. 14-15 (eng), pp. 138-140(15-17) (rus)]. Then $e := \frac{1}{\sqrt{2}}\eta^k$ $(\eta^k := (\eta_0)^k)$. Therefore,

$$\eta_{ij}{}^k := \sum_{I=0}^{n-1} \frac{1}{\sqrt{2}}(\eta_I)_i(g_I)_j{}^k, \tag{6}$$

where $g_{ij} =: (g_0)_{ij}$, and the remaining tensors $(g_I)_{ij}$ are antisymmetric tensors. Indeed,

$$\eta_{(ij)}{}^k := \sqrt{2}(2\eta_{(i}{}^{AB}\eta_{j)(CA)}\eta^k{}_{DB}\theta^{CD} - \eta_{(i}{}^{AB}\eta_{j)AC}\eta^k{}_{DB}\theta^{CD}) = 2(\tfrac{1}{\sqrt{2}}\eta_{(i})\delta_{j)}{}^k - g_{ij}(\tfrac{1}{\sqrt{2}}\eta^k) \tag{7}$$

then the axioms 1-3, the identity (3a) are executed. Analogically,

$$\begin{gathered}\eta_{(i|j|k)} = \sqrt{2}\eta_{(i}{}^{AB}\eta_{|jCA|}\eta_{k)DB}\theta^{CD} = (\tfrac{1}{\sqrt{2}}\eta_j)g_{ik},\\ \eta_{i(jk)} = \sqrt{2}(2\eta_i{}^{AB}\eta_{(j|(CA)|}\eta_{k)DB}\theta^{CD} - \eta_i{}^{AB}\eta_{(j|AC|}\eta_{k)DB}\theta^{CD}) =\\ = \sqrt{2}(\eta_{(j}g_{k)i} - \eta_{(k}g_{j)i} + \tfrac{1}{2}\eta_i g_{jk}) = (\tfrac{1}{\sqrt{2}}\eta_i)g_{jk}\end{gathered} \tag{8}$$

then the identity (3b) is executed,

$$\begin{gathered}\eta_{[ij]}{}^k(\tfrac{1}{\sqrt{2}}\eta_k) = \sqrt{2}\eta_{[i}{}^{AB}\eta_{j]CA}(\tfrac{1}{\sqrt{2}}\eta_k)\eta^k{}_{DB}\theta^{CD} = \eta_{[i}{}^A{}_B\eta_{j]CA}\theta^{BC} =\\ = \underbrace{2\eta_{[i}{}^A{}_B\eta_{j](CA)}\theta^{BC}}_{=\eta_{[i}\eta_{j]}=0} - \underbrace{\eta_{[i}{}^A{}_B\eta_{j]AC}\theta^{(BC)}}_{=0} = 0,\\ < ab - ba, e >=< (a+e)b - b(a+e), a+e > - < ab-ba, a >= 0.\end{gathered} \tag{9}$$

then the identity (3b′) is executed as a consequence from (3b). Indeed ($\varepsilon^{[A|(BC)|D]} = 0$, $\varepsilon^{ABCD} := \eta_i{}^{AB}\eta^{iCD}$ [3, p. 40, Definition 8.2 (eng), p. 169(46), Definition 8.2 (rus)], $\theta^{XY} = \theta^{(XY)}$)

$$\begin{gathered}\tfrac{1}{2}\eta_{(i|j}{}^k\eta_{k|l)}{}^m = \eta_{(i}{}^{AB}\eta_{l)ZC}\eta_k{}^{CD}\eta_{jXA}\eta^k{}_{YB}\eta^m{}_{TD}\theta^{XY}\theta^{ZT} =\\ = \eta_{(i}{}^{CD}\eta_{l)ZC}\eta_j\eta^m{}_{TD}\theta^{ZT} - \varepsilon^{ABCD}\eta_{(l|ZC|}\eta_{i)YB}\eta_{jXA}\eta^m{}_{TD}\theta^{XY}\theta^{ZT} =\\ = \eta_{(i}{}^{CD}\eta_{l)}\eta_j\eta^m{}_{TD}\theta^{ZT}\varepsilon_{ZC} - \tfrac{1}{2}g_{il}\eta_j\eta^m - \varepsilon^{ABCD}\eta_{(l|ZC|}\eta_{i)YB}\eta_{jXA}\eta^m{}_{TD}\theta^{XY}\theta^{ZT},\\[1em] \tfrac{1}{2}\eta_{(i|k}{}^m\eta_{j|l)}{}^k = \eta_{(i}{}^{AB}\eta_{l)ZC}\eta_j{}^{CD}\eta_{kXA}\eta^k{}_{TD}\eta^m{}_{YB}\theta^{XY}\theta^{ZT} =\\ = \eta_{(i}\eta_{l)ZC}\eta_j{}^{CD}\eta^m{}_{TD}\theta^{ZT} - \eta_{(i|TD|}\eta_{l)ZC}\theta^{ZT}\eta_j{}^{CD}\eta^m +\\ + \eta_k{}^{BA}\eta_j{}^{CD}\eta_{(i|XA|}\eta_{l)ZC}\eta^k{}_{TD}\eta^m{}_{YB}\theta^{XY}\theta^{ZT} =\\ = \underbrace{\eta_{(i}\eta_{l)ZC}\eta_j{}^{CD}\eta^m{}_{TD}\theta^{ZT}}_{1} - \underbrace{\eta_{(i|TD|}\eta_{l)ZC}\theta^{ZT}\eta_j{}^{CD}\eta^m}_{2} + \underbrace{\eta_j{}^{BA}\eta_{(i|XA|}\eta_{l)}\eta^m{}_{YB}\theta^{XY}}_{1} -\\ -\varepsilon^{BACD}\eta_{(i|XA|}\eta_{l)ZC}\eta_{jTD}\eta^m{}_{YB}\theta^{XY}\theta^{ZT} =\\[1em] = \underbrace{\eta_{(i}{}^{CD}\eta_{l)}\eta_j\eta^m{}_{TD}\theta^{ZT}\varepsilon_{ZC}}_{1} - \underbrace{\frac{1}{2}g_{il}\eta_j\eta^m}_{2} - \varepsilon^{ABCD}\eta_{(l|ZC|}\eta_{i)YB}\eta_{jTD}\eta^m{}_{XA}\theta^{ZT}\theta^{XY}.\end{gathered} \tag{10}$$

Otherwise, from (3)

$$\begin{gathered}\tfrac{1}{2}((ab)a + a(ab)) =\\ = -\underbrace{< a, (ab) >}_{=<b,e><a,a>}e + < a, e > (ab) + \underbrace{< (ab), e >}_{-<a,b>+2<b,e><a,e>} a =\\ = - < a, b > a + < a, e > (ab) + < b, e > \underbrace{(- < a, a > e + 2 < a, e > a)}_{a^2} =\\ = a\underbrace{(- < a, b > e + < a, e > b + < b, e > a)}_{\frac{1}{2}(ab+ba)},\end{gathered} \tag{10′}$$

$$(ab)a = a(ba).$$

There is the alternative-elastic identity

$$((aa)b) - (a(ab)) - (a(ba)) = ((b(aa)) - ((ba)a) - ((ab)a). \tag{11}$$

$$\begin{gathered}
((aa)b) - (a(ab)) - (a(ba)) = - < a, a > b + 2 < a, e > (ab) - \\
-2(a(- < a, b > e + < a, e > b + < b, e > a)) = \\
= - < a, a > b + 2 < a, b > a - 2 < b, e > (- < a, a > e + 2 < a, e > a) = \\
= 2 < b, e >< a, a > e + (2 < a, b > -4 < b, e >< a, e >)a - < a, a > b, \\
\\
(b(aa)) - ((ba)a)) - ((ab)a) = - < a, a > b + 2 < a, e > (ba) - \\
-2((- < a, b > e + < a, e > b + < b, e > a)a) = \\
= - < a, a > b + 2 < a, b > a - 2 < b, e > (- < a, a > e + 2 < a, e > a) = \\
= 2 < b, e >< a, a > e + (2 < a, b > -4 < b, e >< a, e >)a - < a, a > b,
\end{gathered} \tag{12}$$

then the axioms 4-5 are executed too. Note that the equations (3) provide the execution of the axiom 3-5, the equations (5),(4) provide the execution of the axiom 1-2,6 and the equations (3). From (3), the common Jordan identity

$$a^k(ba^l) = (a^k b)a^l \tag{13}$$

follows. □

In addition,

$$g_{kr}\eta_{(i|j|}{}^k\eta_{l)m}{}^r = g_{kr}\eta^k{}_{j(i}\eta^r{}_{|m|l)}. \tag{14}$$

This identity follows from (3) and it is called **weakly normalization identity**. This identity is the normalization identity for n=8 only ($g_{kr}\eta^k{}_{j(i}\eta^r{}_{|m|l)} = g_{jm}g_{il}$ in this case). And so this algebra is normalized [3, p. 78, eq. (14.97) (eng), p. 212(89), eq. (14.97) (rus)].

**Theorem 2.** *The metric real numbers, complex numbers, quaternions, octonions, sedenions, hypercomplex Cayley-Dickson numbers possess the identities (3).*

*Proof.* Let $r = \overline{1, n-1}$ then for the Euclidean metric $\delta_{ij}$, $\forall x$

$$\begin{gathered}
x = x_0 e + \sum_{r=1}^{n-1} i_r x_r, \quad \bar{x} = x_0 e - \sum_{r=1}^{n-1} i_r x_r, \\
i_r i_s = -i_s i_r, \quad i_r i_r = -e, \quad e i_r = i_r e = i_r, \quad ee = e, \quad < i_r, e >= 0.
\end{gathered} \tag{15}$$

Let $x := a + bi$, $y := c + di$, where $i := i_{n/2}$ $(r = \overline{1, n/2 - 1})$ then for the Euclidean metric $\delta_{ij}$, $\forall a, b$

$$a = a_0 e + \sum_{r=1}^{n/2-1} a_r i_r, \quad ie = i, \quad i_r i = i_{r+n/2}, \quad < e, ai >= 0. \tag{16}$$

**Definition 2.** *(according to [6, pp.300-303]) 1. Let's define the multiplication for an inductive step according to the Cayley-Dickson double procedure as*

$$a(bi) = (ba)i, \quad (ai)b = (a\bar{b})i, \quad (ai)(bi) = -\bar{b}a. \tag{17}$$

*2. Let's define the conjugation for an inductive step according to the Cayley-Dickson double procedure as*

$$\overline{a + bi} = \bar{a} - bi. \tag{18}$$

The metric hypercomplex Cayley-Dickson algebra possesses the following identities.
1.
Set $\forall a, c: \ \frac{1}{2}(a\bar{c} + c\bar{a}) =< a, c > e$ by the induction hypothesis then the following identity is obtained

$$\begin{array}{c} \frac{1}{2}(x\bar{y} + y\bar{x}) =< x, y > e, \\ (x\bar{y} + y\bar{x}) = (a + bi)(\bar{c} - di) + (c + di)(\bar{a} - bi) = \\ = a\bar{c} + \bar{d}b + (bc - da)i + c\bar{a} + \bar{b}d - (bc - da)i = \\ = 2(< a, c > e + < d, b > e) = 2 < x, y > e. \end{array} \tag{19}$$

2.

$$\begin{array}{c} \frac{1}{2}(yx + xy) = -\frac{1}{2}(y\bar{x} + x\bar{y}) + < x, e > y + < y, e > x = \\ = - < x, y > e + < x, e > y + < y, e > x. \end{array} \tag{20}$$

3.

$$2 < ai, b >= (ai)\bar{b} + b(\overline{ai}) = (ai)\bar{b} - b(ai) = (ab)i - (ab)i = 0. \tag{21}$$

4.
Set $\forall a, c: \ \overline{ac} = \bar{c}\bar{a}$ by the induction hypothesis then the following identity is obtained

$$\begin{array}{c} \overline{xy} = \overline{(a + bi)(c + di)} = \overline{ac - \bar{d}b + (b\bar{c} + da)i} = \\ = (\bar{c}\bar{a} - \bar{b}d) - (b\bar{c} + da)i = \bar{y}\bar{x}. \end{array} \tag{22}$$

5.
Set $\forall a, c : < ac - ca, e >= 0$ by the induction hypothesis then the following identity is obtained

$$\begin{array}{c} < xy - yx, e >= \\ =< ((ac - \bar{d}b) + (b\bar{c} + da)i) - ((ca - \bar{b}d) + (bc + d\bar{a})i), e >= \\ =< (ac - ca) + ((-b + 2 < b, e >)d - (-d + 2 < d, e >)b, e >= 0. \end{array} \tag{23}$$

6.
Set $\forall a, c : < ac, a >=< ca, a >=< a\bar{c}, a >=< \bar{c}a, a >=< c, e >< a, a >$ by the induction hypothesis then the following identity is obtained

$$\begin{array}{c} < xy, x >=< y, e >< x, x >, \\ < xy, x >= \\ =< ac - \bar{d}b + (b\bar{c} + da)i, a + bi >= \\ =< ac, a > + < bc, b > + \underbrace{< da, b > + < db, a > -2 < d, e >< b, a >}_{=-<d(a-b),(a-b)>+<d,e><a,a>+<d,e><b,b>-2<d,e><b,a>=0} = \\ =< y, e >< x, x > . \end{array} \tag{24}$$

□

On the other hand, on the base of Corollary 8.2 [3, p.44 (eng),p. 174(51) (rus)],

the relations

$$\eta_{ij}{}^k := \sum_{I=0}^{n-1}(\tfrac{1}{\sqrt{2}}(\eta_I)^k)(-3(h_I)_{ij} + \eta_{(j}(\eta_I)_{i)}) =$$
$$= \sqrt{2}\eta_i{}^{AB}\eta_{jCA}\eta^k{}_{DB}\tfrac{2}{N}\varepsilon^{CD} - 3\sum_{I=1}^{n-1}(\tfrac{1}{\sqrt{2}}(\eta_I)^k)(h_I)_{ij} =$$
$$= \sqrt{2}(2-n)\eta_i{}^{AB}\eta_{jCA}\eta^k{}_{DB}\tfrac{2}{N}\varepsilon^{CD} + \sum_{I=1}^{n-1}(-3(\tfrac{1}{\sqrt{2}}(\eta_I)^k)(h_I)_{ij} + \sqrt{2}\eta_i{}^{AB}\eta_{jCA}\eta^k{}_{DB}\tfrac{2}{N}\varepsilon^{CD}) \tag{25}$$

are executed, where $g_{ij} =: 3(h_0)_{ij}$, and the remaining tensors $(h_I)_{ij}$ are antisymmetric tensors. But $(h_I)_{ij}$ are not arbitrary tensors, there are the compatibility conditions (3)

$$\begin{array}{ll} \eta_{(i|j|k)} = (\tfrac{1}{\sqrt{2}}\eta_j)g_{ik}, & \eta_{i(jk)} = (\tfrac{1}{\sqrt{2}}\eta_i)g_{jk}, \\ \sum_{I=1}^{n-1}((\tfrac{1}{\sqrt{2}}(\eta_I)_{(k})(h_{|I|})_{i)j} = 0, & \sum_{I=1}^{n-1}((\tfrac{1}{\sqrt{2}}(\eta_I)_{(k})(h_{|I|})_{j)i} = 0. \end{array} \tag{26}$$

So the equation (25) takes the form

$$\eta_{ij}{}^k := \sum_{I=0}^{n-1}(\tfrac{1}{\sqrt{2}}(\eta_I)^k)(-3(h_I)_{ij} + \eta_{(j}(\eta_I)_{i)}) =$$
$$= \sqrt{2}(2-n)\eta_i{}^{AB}\eta_{jCA}\eta^k{}_{DB}\tfrac{2}{N}\varepsilon^{CD} +$$
$$+ \sum_{I=1}^{n-1}(\underbrace{((\frac{1}{\sqrt{2}}(\eta_I)_j)(h_I)_i{}^k - (\frac{1}{\sqrt{2}}(\eta_I)_i)(h_I)_j{}^k - (\frac{1}{\sqrt{2}}(\eta_I)^k)(h_I)_{ij})}_{:=(h_I)_{ij}{}^k} + \sqrt{2}\eta_i{}^{AB}\eta_{jCA}\eta^k{}_{DB}\tfrac{2}{N}\varepsilon^{CD}). \tag{27}$$

**Note 1.** *Since for any special (non-special) orthogonal transformation $S_i^j$, according to Corollary 8.3 [3, p. 44 (eng), p. 174(51) (rus)], the equation*

$$S_i^j\eta_j{}^{AB} = \eta_i{}^{CD}\tilde{S}_C{}^A\tilde{\tilde{S}}_D{}^B \ (S_i^j\eta_j{}^{AB} = \eta_i{}^{DC}\tilde{S}_C{}^A\tilde{\tilde{S}}_D{}^B) \tag{28}$$

*is executed then any special (non-special) orthogonal transformation $S_i^j$ keeping the algebra identity ($S := \tilde{S} = \tilde{\tilde{S}}$) will transform the structural constants as $S_l{}^iS_m{}^j\eta_{ij}{}^kS^r{}_k$ that generates the transformation of the controlling spin-tensor $\theta^{AB} \longmapsto \theta^{CD}S^A{}_CS^B{}_D$, ($\theta^{AB} \longmapsto \frac{4}{N}\varepsilon^{AB} - \theta^{CD}S^A{}_CS^B{}_D$) keeping without a change $\eta_i{}^{CD}$ from (5).*

**Definition 3.** *Hypercomplex orthogonal algebra $\mathbb{A}$ is called* `homogenous algebra` *if the orthogonal transformations $S_I$ exist for all $I$: $(h_I)_{ij} = \alpha_I(S_I)_i{}^m(h_{gen})_{ml}(S_I)_j{}^l$, $(\eta_I)_i = (S_I)_i{}^m(\eta_{gen})_m$ ($\alpha_I \in \mathbb{R}$, $I = \overline{1, n-1}$).*

So, in order to construct an hypercomplex orthogonal homogenous algebra $\mathbb{A}$, the algebra identity $\frac{1}{\sqrt{2}}\eta^k$ and the generator $\frac{1}{\sqrt{2}}(\eta_{gen})^k(h_{gen})_{ij}$ are necessary to know. Then using orthogonal transformations keeping the algebra identity, the $n-1$ basic elements (27) are constructed from this generator.

$$\eta_{ij}{}^k = \overbrace{(1 - \sum_{I=1}^{n-1}\alpha_I)}^{:=\alpha_0}\underbrace{\sqrt{2}\eta_i{}^{AB}\eta_{jCA}\eta^k{}_{DB}\overbrace{\frac{2}{N}\varepsilon^{CD}}^{:=(\theta_0)^{CD}}}_{:=(\eta_0)_{ij}{}^k} -$$
$$+ \sum_{I=1}^{n-1}\alpha_I\underbrace{((S_I)_i{}^l(h_{gen})_{lm}{}^r(S_I)^k{}_r(S_I)_j{}^m + \sqrt{2}\eta_i{}^{AB}\eta_{jCA}\eta^k{}_{DB}\frac{2}{N}\varepsilon^{CD})}_{:=(\eta_I)_{ij}{}^k} \tag{29}$$

Obviously, the equation (29) is a decomposition of the hypercomplex orthogonal homogenous algebra $\mathbb{A}\ ({\eta_{ij}}^k)$ on an algebraic basis $\mathbb{A}_I\ ({(\eta_I)_{ij}}^k)$. In other way, ${(\eta_I)_{ij}}^k := \sqrt{2}{\eta_i}^{AB}\eta_{jCA}\,{\eta^k}_{DB}(\theta_I)^{CD}$. Define $\theta^{CD} := \sum\limits_{I=0}^{n-1}\alpha_I(\theta_I)^{CD}$ then ${\eta_{ij}}^k$ determine the hypercomplex orthogonal homogenous algebra $\mathbb{A}$ according to (5).

Consider Algorithm 9.1 [3, pp. 50-52 (eng), pp. 181-182(58-59) (rus)] based on the Bott periodicity [4]:

**Algorithm 1.** *1.* $\Lambda, ... = \overline{1,n},\ i, ... = \overline{1,n},\ A, ... = \overline{1, 2^{\frac{n}{2}-1}},\ \alpha, ... = \overline{1, n+6},\ a, ... = \overline{1, 2^{\frac{n+6}{2}-1}}$. *Suppose there is an orthogonal algebra* $\mathbb{A}$ *with the structural constants generated from the connecting operators* ${\eta_\Lambda}^{AB}$, *the metric spinor* $\varepsilon^{XZ}$, *and the inclusion operator* ${H_i}^\Lambda$. *We assume that the metric tensor* $g_{\Lambda\Psi}$ *on the main diagonal contains «+» only. Then we can construct the antisymmetric operators for the space* $\mathbb{C}^{n+6}$

$$
{\eta_\alpha}^{ab} = -{\eta_\alpha}^{ba} :=
$$
$$
\begin{pmatrix}
0 & 0 & 0 & \xi\varepsilon^{AQ} & 0 & \gamma\varepsilon^{AK} & -\alpha\varepsilon^{AD} & {\eta_\Lambda}^A{}_B \\
0 & 0 & -\xi\varepsilon_{CR} & 0 & -\gamma\varepsilon_{CM} & 0 & {(\eta^T)_{\Lambda C}}^D & \beta\varepsilon_{CB} \\
0 & \xi\varepsilon_{NY} & 0 & 0 & -\alpha\varepsilon_{NM} & {(\eta^T)_{\Lambda N}}^K & 0 & \delta\varepsilon_{NB} \\
-\xi\varepsilon^{LZ} & 0 & 0 & 0 & {\eta_\Lambda}^L{}_M & \beta\varepsilon^{LK} & -\delta\varepsilon^{LD} & 0 \\
0 & \gamma\varepsilon_{PY} & \alpha\varepsilon_{PR} & -{(\eta^T)_{\Lambda P}}^Q & 0 & 0 & 0 & -\zeta\varepsilon_{PB} \\
-\gamma\varepsilon^{SZ} & 0 & -{\eta_\Lambda}^S{}_R & -\beta\varepsilon^{SQ} & 0 & 0 & \zeta\varepsilon^{SD} & 0 \\
\alpha\varepsilon^{XZ} & -{\eta_\Lambda}^X{}_Y & 0 & \delta\varepsilon^{XQ} & 0 & -\zeta\varepsilon^{XK} & 0 & 0 \\
-{(\eta^T)_{\Lambda T}}^Z & -\beta\varepsilon_{TY} & -\delta\varepsilon_{TR} & 0 & \zeta\varepsilon_{TM} & 0 & 0 & 0
\end{pmatrix}, \tag{30}
$$

$$
\eta_{\alpha ab} = -\eta_{\alpha ba} :=
$$
$$
\begin{pmatrix}
0 & 0 & 0 & -\zeta\varepsilon_{AQ} & 0 & -\delta\varepsilon_{AK} & -\beta\varepsilon_{AD} & {\eta_{\Lambda A}}^B \\
0 & 0 & \zeta\varepsilon^{CR} & 0 & \delta\varepsilon^{CM} & 0 & (\eta^T)_\Lambda{}^C{}_D & \alpha\varepsilon^{CB} \\
0 & -\zeta\varepsilon^{NY} & 0 & 0 & -\beta\varepsilon^{NM} & (\eta^T)_\Lambda{}^N{}_K & 0 & -\gamma\varepsilon^{NB} \\
\zeta\varepsilon_{LZ} & 0 & 0 & 0 & {\eta_{\Lambda L}}^M & \alpha\varepsilon_{LK} & \gamma\varepsilon_{LD} & 0 \\
0 & -\delta\varepsilon^{PY} & \beta\varepsilon^{PR} & -(\eta^T)_\Lambda{}^P{}_Q & 0 & 0 & 0 & \xi\varepsilon^{PB} \\
\delta\varepsilon_{SZ} & 0 & -{\eta_{\Lambda S}}^R & -\alpha\varepsilon_{SQ} & 0 & 0 & -\xi\varepsilon_{SD} & 0 \\
\beta\varepsilon_{XZ} & -{\eta_{\Lambda X}}^Y & 0 & -\gamma\varepsilon_{XQ} & 0 & \xi\varepsilon_{XK} & 0 & 0 \\
-(\eta^T)_\Lambda{}^T{}_Z & -\alpha\varepsilon^{TY} & \gamma\varepsilon^{TR} & 0 & -\xi\varepsilon^{TM} & 0 & 0 & 0
\end{pmatrix}, \tag{30'}
$$

$$
\begin{aligned}
\alpha &:= \tfrac{1}{2}(i\eta_{n+1} + \eta_{n+2}), \quad \beta := \tfrac{1}{2}(-i\eta_{n+1} + \eta_{n+2}), \\
\gamma &:= \tfrac{1}{2}(\eta_{n+3} + i\eta_{n+4}), \quad \delta := \tfrac{1}{2}(-\eta_{n+3} + i\eta_{n+4}), \\
\xi &:= \tfrac{1}{2}(\eta_{n+5} + i\eta_{n+6}), \quad \zeta := \tfrac{1}{2}(-\eta_{n+5} + i\eta_{n+6}).
\end{aligned} \tag{31}
$$

*2.* $\Lambda, ... = \overline{1, n+8},\ i, ... = \overline{1, n+8},\ A, ... = \overline{1, 2^{\frac{n+8}{2}-1}},\ \alpha, ... = \overline{1, n+6},\ a, ... = \overline{1, 2^{\frac{n+6}{2}-1}}$. *Transition to the connecting operators of the space* $\mathbb{C}^{n+8}$ *is carried out as follows:*

$$
{\eta_\Lambda}^{AB} := \begin{pmatrix} {\eta_\alpha}^{ab} & \phi{\delta^a}_d \\ \psi{\delta_c}^b & -(\eta^T)_{\alpha cd} \end{pmatrix}, \tag{32}
$$

$$
\phi := \frac{1}{2}(i\eta_{n+7} + \eta_{n+8}), \quad \psi := \frac{1}{2}(-i\eta_{n+7} + \eta_{n+8}) \tag{33}
$$

*with the metric spinor* $\varepsilon^{XZ} := \begin{pmatrix} 0 & {\delta^a}_d \\ {\delta_c}^b & 0 \end{pmatrix}$, $\varepsilon_{XZ} := \begin{pmatrix} 0 & {\delta_a}^d \\ {\delta^c}_b & 0 \end{pmatrix}$. *Then we go to the connecting operators of the space* $\mathbb{R}^{n+8} \subset \mathbb{C}^{n+8}$ *using the corresponding* *inclusion operator. And such the operators generate the structure constants of the hypercomplex algebra with dimension equal to $n+8$.*

**Note 2.** *In the conditions of Algorithm 1 and Examples 1,2, the algebra identity is $\frac{1}{\sqrt{2}}\eta_{n+8}$ (or accordingly $\frac{1}{\sqrt{2}}\eta_n$). Therefore, for reduction of designations in conformity, it is necessary to make the redesignation: $n+8 \longmapsto 0$ (or accordingly $n \longmapsto 0$).*

**Example 1.** *Let $\mathbb{A}_{gen}$ be defined with the help of the condition on the controlling spinor*

$$X^1 := 1, \quad X^{2^{\frac{n+8}{2}-2}+1} := 1. \tag{34}$$

*Let $\| H_{i_{n+8}}{}^{\Lambda_{n+8}} \|$ ($i_{n+8}, ... = \overline{1, n+8},\ A_{n+8}, ... = \overline{1, 2^{\frac{n+8}{2}-1}}$) be identity matrix. Then*

$$(\theta_{gen})^{C_{n+8}D_{n+8}} = X^{C_{n+8}}X^{D_{n+8}}, \tag{35}$$

$$\eta_{i_{n+8}j_{n+8}}{}^{k_{n+8}} := \sqrt{2}\eta_{i_{n+8}}{}^{A_{n+8}B_{n+8}} \underbrace{\eta_{j_{n+8}C_{n+8}A_{n+8}}X^{C_{n+8}}}_{:=P_{j_{n+8}A_{n+8}}} \underbrace{\eta^{k_{n+8}}{}_{D_{n+8}B_{n+8}}X^{D_{n+8}}}_{:=P^{k_{n+8}}{}_{B_{n+8}}}. \tag{36}$$

*Therefor, ($i_n, ... = \overline{1, n},\ A_n, ... = \overline{1, 2^{\frac{n}{2}-1}}$)*

$$X^{A_{n+8}} = (X^{B_n}, 0, 0, 0, 0, 0, 0, 0, Y_{C_n}, 0, 0, 0, 0, 0, 0, 0), \quad X^1 = 1, \quad Y_1 = 1, \tag{37}$$

$$P_{j_{n+8}A_{n+8}} := H_{j_{n+8}}{}^{\Lambda_{n+8}}(\phi Y_{A_n}, 0, 0, -\zeta X_{B_n}, 0, -\delta X_{C_n}, -\beta X_{D_n}, \eta_{\Lambda_n K_n}{}^{M_n}X^{K_n}, \\ \psi X^{L_n}, 0, 0, \xi Y^{P_n}, 0, \gamma Y^{Q_n}, -\alpha Y^{R_n}, \eta_{\Lambda_n}{}^{Z_n}{}_{T_n}Y_{Z_n}). \tag{38}$$

*Define*

$$h_{i_{n+8}j_{n+8}} := H_{i_{n+8}}{}^{\Lambda_{n+8}}H_{j_{n+8}}{}^{\Psi_{n+8}}(\phi\psi - \xi\zeta - \gamma\delta + \alpha\beta + \eta_{\Lambda_n}{}^{A_n}{}_{B_n}Y_{A_n}\eta_{\Psi_n C_n}{}^{B_n}X^{C_n}). \tag{39}$$

*Then*

$$g_{i_{n+8}j_{n+8}} = 2h_{(i_{n+8}j_{n+8})}, \quad (h_{gen})_{i_{n+8}j_{n+8}} := 2ih_{[i_{n+8}j_{n+8}]}. \tag{40}$$

*Whence, ($i, ... = \overline{1, n+8},\ A, ... = \overline{1, 2^{\frac{n+8}{2}-1}}$) for all n mod 8 = 0*

Table 1: The matrix table example of $(h_{gen})_{ij}$.

| | | | | | | | | | | | | | | | | |
|---|---|---|---|---|---|---|---|---|---|---|---|---|---|---|---|---|
| | -1 | | | | | | | ⋯ | | | | | | | | |
| 1 | | | | | | | | ⋯ | | | | | | | | |
| | | | -1 | | | | | ⋯ | | | | | | | | |
| | | 1 | | | | | | ⋯ | | | | | | | | |
| | | | | | -1 | | | ⋯ | | | | | | | | |
| | | | | 1 | | | | ⋯ | | | | | | | | |
| | | | | | | | 1 | ⋯ | | | | | | | | |
| | | | | | | -1 | | ⋯ | | | | | | | | |
| ⋮ | ⋮ | ⋮ | ⋮ | ⋮ | ⋮ | ⋮ | ⋮ | ⋱ | ⋮ | ⋮ | ⋮ | ⋮ | ⋮ | ⋮ | ⋮ | ⋮ |
| | | | | | | | | ⋯ | | 1 | | | | | | |
| | | | | | | | | ⋯ | -1 | | | | | | | |
| | | | | | | | | ⋯ | | | | -1 | | | | |
| | | | | | | | | ⋯ | | | 1 | | | | | |
| | | | | | | | | ⋯ | | | | | | -1 | | |
| | | | | | | | | ⋯ | | | | | 1 | | | |
| | | | | | | | | ⋯ | | | | | | | | |
| | | | | | | | | ⋯ | | | | | | | | |

Table 2: The multiplication table example of the algebra $\mathbb{A}_{gen}$ $((\eta_{gen})_{ij}{}^{k})$.

| * | $e_1$ | $e_2$ | $e_3$ | $e_4$ | $e_5$ | $e_6$ | $e_7$ | $e_8$ | $\cdots$ | $e_{n+1}$ | $e_{n+2}$ | $e_{n+3}$ | $e_{n+4}$ | $e_{n+5}$ | $e_{n+6}$ | $e_{n+7}$ | $e_{n+8}$ |
|---|---|---|---|---|---|---|---|---|---|---|---|---|---|---|---|---|---|
| $e_1$ | $-e_{n+8}$ | $e_{n+7}$ | | | | | | | $\cdots$ | | | | | | | $-e_2$ | $e_1$ |
| $e_2$ | $-e_{n+7}$ | $-e_{n+8}$ | | | | | | | $\cdots$ | | | | | | | $e_1$ | $e_2$ |
| $e_3$ | | | $-e_{n+8}$ | $e_{n+7}$ | | | | | $\cdots$ | | | | | | | $-e_4$ | $e_3$ |
| $e_4$ | | | $-e_{n+7}$ | $-e_{n+8}$ | | | | | $\cdots$ | | | | | | | $e_3$ | $e_4$ |
| $e_5$ | | | | | $-e_{n+8}$ | $e_{n+7}$ | | | $\cdots$ | | | | | | | $-e_6$ | $e_5$ |
| $e_6$ | | | | | $-e_{n+7}$ | $-e_{n+8}$ | | | $\cdots$ | | | | | | | $e_5$ | $e_6$ |
| $e_7$ | | | | | | | $-e_{n+8}$ | $-e_{n+7}$ | $\cdots$ | | | | | | | $e_8$ | $e_7$ |
| $e_8$ | | | | | | | $e_{n+7}$ | $-e_{n+8}$ | $\cdots$ | | | | | | | $-e_7$ | $e_8$ |
| $\vdots$ | $\vdots$ | $\vdots$ | $\vdots$ | $\vdots$ | $\vdots$ | $\vdots$ | $\vdots$ | $\vdots$ | $\ddots$ | $\vdots$ | $\vdots$ | $\vdots$ | $\vdots$ | $\vdots$ | $\vdots$ | $\vdots$ | $\vdots$ |
| $e_{n+1}$ | | | | | | | | | $\cdots$ | $-e_{n+8}$ | $-e_{n+7}$ | | | | | $e_{n+2}$ | $e_{n+1}$ |
| $e_{n+2}$ | | | | | | | | | $\cdots$ | $e_{n+7}$ | $-e_{n+8}$ | | | | | $-e_{n+1}$ | $e_{n+2}$ |
| $e_{n+3}$ | | | | | | | | | $\cdots$ | | | $-e_{n+8}$ | $e_{n+7}$ | | | $-e_{n+4}$ | $e_{n+3}$ |
| $e_{n+4}$ | | | | | | | | | $\cdots$ | | | $-e_{n+7}$ | $-e_{n+8}$ | | | $e_{n+3}$ | $e_{n+4}$ |
| $e_{n+5}$ | | | | | | | | | $\cdots$ | | | | | $-e_{n+8}$ | $e_{n+7}$ | $-e_{n+6}$ | $e_{n+5}$ |
| $e_{n+6}$ | | | | | | | | | $\cdots$ | | | | | $-e_{n+7}$ | $-e_{n+8}$ | $e_{n+5}$ | $e_{n+6}$ |
| $e_{n+7}$ | $e_2$ | $-e_1$ | $e_4$ | $-e_3$ | $e_6$ | $-e_5$ | $-e_8$ | $e_7$ | $\cdots$ | $-e_{n+2}$ | $e_{n+1}$ | $e_{n+4}$ | $-e_{n+3}$ | $e_{n+6}$ | $-e_{n+5}$ | $-e_{n+8}$ | $e_{n+7}$ |
| $e_{n+8}$ | $e_1$ | $e_2$ | $e_3$ | $e_4$ | $e_5$ | $e_6$ | $e_7$ | $e_8$ | $\cdots$ | $e_{n+1}$ | $e_{n+2}$ | $e_{n+3}$ | $e_{n+4}$ | $e_{n+5}$ | $e_{n+6}$ | $e_{n+7}$ | $e_{n+8}$ |

$$\begin{aligned}(\eta_{gen})_{ij}{}^{k} := (\tfrac{1}{\sqrt{2}}(\eta_{n+8})_i)\delta_j{}^k + (\tfrac{1}{\sqrt{2}}(\eta_{n+8})_j)\delta_i{}^k - g_{ij}(\tfrac{1}{\sqrt{2}}(\eta_{n+8})^k)+\\ +(\tfrac{1}{\sqrt{2}}(\eta_{n+7})_j)(h_{gen})_i{}^k - (\tfrac{1}{\sqrt{2}}(\eta_{n+7})_i)(h_{gen})_j{}^k - (h_{gen})_{ij}(\tfrac{1}{\sqrt{2}}(\eta_{n+7})^k).\end{aligned} \tag{41}$$

*Thus, the generating algebra $\mathbb{A}_{gen}(h_{gen})$ is constructed. This algebra is unique in that, it is the generator for the metric Cayley-Dickson algebra for $n+8=2^k$. And because the metric Cayley-Dickson algebra with $\eta_{ij}{}^k$ satisfies the equations (3) and Definition 3 then it must have the generating algebra with, for example, $(\tilde{h}_{gen})_{ij} := \eta_{ij}{}^k(\frac{1}{\sqrt{2}}\eta_{n+7})_k$, $\tilde{h}_{gen} = \alpha S h_{gen} S$ for some $\alpha \in \mathbb{R}$, where $S$ is the orthogonal transformation keeping the algebra identity.*

**Note 3.** *The controlling spinor $X^A = (1,0,0,0,1,0,0,0)$ from (37) generates the octonion algebra entirely. But tensor*

$$h_{ij} := H_i{}^{\Lambda} H_j{}^{\Psi}(\eta_{\Lambda}{}^{A}{}_{B} Y_A \eta_{\Psi C}{}^{B} X^C),\ \ X^C = (1,0,0,0,0,0,0,0),\ \ Y_A = (1,0,0,0,0,0,0,0) \tag{42}$$

*is generated by means of the equation (39) as usual. Therefore, $n=8$ is the initial induction step.*

The last paragraph of the example 1 should be clarified. Let the hypercomplex orthogonal algebra be given by the structural constants $\eta_{ij}{}^k$. Then these constants can be expanded according to (3) as

$$\eta_{ij}{}^k := (\eta_0)_{ij}{}^k + (\eta_a)_{ij}{}^k. \tag{43}$$

From (29) and Corollary 8.2 [3, p. 44 (eng), p. 174(51) (rus)]

$$\begin{gathered}(\eta_0)_{ij}{}^k := (\tfrac{1}{\sqrt{2}}\eta_i)\delta_j{}^k + (\tfrac{1}{\sqrt{2}}\eta_j)\delta_i{}^k - g_{ij}(\tfrac{1}{\sqrt{2}}\eta^k) = \sqrt{2}\eta_i{}^{AB}\eta_{jCA}\eta^k{}_{DB}\underbrace{\frac{2}{N}\varepsilon^{CD}}_{(\theta_0)^{CD}},\\ \eta_{(ij)k} = (\eta_0)_{(ij)k},\quad \eta_{(i|j|k)} = (\eta_0)_{(i|j|k)},\quad \eta_{i(jk)} = (\eta_0)_{i(jk)},\\ \begin{cases}(\eta_a)_{(ij)k} = 0,\\ (\eta_a)_{(i|j|k)} = 0,\\ (\eta_a)_{i(jk)} = 0,\end{cases} \Rightarrow (\eta_a)_{ijk} = \eta_{[ijk]},\quad (\tfrac{1}{\sqrt{2}}\eta^i)(\eta_a)_{ijk} = 0.\end{gathered} \tag{44}$$

Define

$$(\eta_a)_{ij}{}^k =: \sqrt{2}\eta_i{}^{AB}\eta_{jCA}\eta^k{}_{DB}(\theta_a)^{CD} = -\sqrt{2}\eta_i{}^{AB}\eta_{jAC}\eta^k{}_{DB}(\theta_a)^{CD}. \tag{45}$$

This equation always has the particular solution (see [3, eq. (12.19), (12.20), p. 63 (eng), p. 195(72) (rus)])

$$(\theta_a)^{CD} = (\theta_a)^{DC} := -\frac{4}{3\sqrt{2}N}(\eta_a)_{lm}{}^r\eta^l{}_{XY}\eta^{mXC}\eta_r{}^{DY} = \frac{4}{3\sqrt{2}N}(\eta_a)_{lm}{}^r\eta^l{}_{XY}\eta^{mCX}\eta_r{}^{DY}. \tag{46}$$

This statement follows from the equation executed for all even $n \geq 8$ (Note 16.1 [3, p. 107 (eng), p. 244(121) (rus)])

$$\eta_{[i}{}^{AB}\eta_{j]AC}\eta^{[m|XC|}\eta^{l]}{}_{XY}\eta_r{}^{DY}\eta^k{}_{DB} = \tfrac{N}{2}(\delta_r{}^{[l}\delta_{[j}{}^{m]}\delta_{i]}{}^k - g^{k[l}\delta_{[j}{}^{m]}g_{i]r}) + \tfrac{N}{4}\delta_r{}^k\delta_i{}^{[l}\delta_j{}^{m]}. \tag{47}$$

And this identity is a consequence of the identities (16.28) and (16.31) [3, p. 105 (eng), pp. 241-242(118-119) (rus)]. Thus,

$$\theta^{CD} := (\theta_0)^{CD} + (\theta_a)^{CD} \tag{48}$$

that proves the theorem.

**Theorem 3.** *Every hypercomplex orthogonal algebra $\mathbb{A}$ admits the decomposition (5).*

**Theorem 4.** *Hypercomplex metric Cayley-Dickson algebra is* `hypercomplex special orthogonal homogenous algebra` *$\mathbb{A}$.*

*Proof.* Let $x := a + bi$, where $i := i_{n/2}$. Set that any hypercomplex metric Cayley-Dickson algebra $\mathbb{A}^{\frac{n}{2}}$ is the hypercomplex special orthogonal homogenous algebra by the induction hypothesis. Then $(h_{gen})_{ij} := (h_1)_{ij}$ and $(h_I)_{ij} = \alpha_I(S_I)_i{}^m(h_{gen})_{ml}(S_I)_j{}^l$ ($\alpha_I \in \mathbb{R},\ I = \overline{1, \frac{n}{2}-1},\ i,j,m,l = \overline{0, \frac{n}{2}-1},\ a,b,c = \overline{\frac{n}{2}, n-1},\ \alpha,\beta = \overline{0, n-1}$). Let $h_{gen}$ have the form ($h_{ab} := \delta_a{}^i\delta_b{}^j h_{ij},\ h_{aj} := \delta_a{}^i h_{ij},\ h_{ib} := \delta_b{}^j h_{ij},\ \delta_a{}^i : i_{I+n/2} \to i_I$))

$$(h_{gen})_{\alpha\beta} = \begin{pmatrix} (h_1)_{ij} & 0 \\ 0 & -(h_1)_{ab} + \frac{1}{2}((\eta_{n/2})_a(\eta_{1+n/2})_b - (\eta_{1+n/2})_a(\eta_{n/2})_b) \end{pmatrix}. \tag{49}$$

Then for the hypercomplex metric Cayley-Dickson algebra $\mathbb{A}^n$ only the three types of the basic elements exist.

1.

$$(h_I)_{\alpha\beta} = \begin{pmatrix} (h_I)_{ij} & 0 \\ 0 & -(h_I)_{ab} + \frac{1}{2}((\eta_{n/2})_a(\eta_{I+n/2})_b - (\eta_{I+n/2})_a(\eta_{n/2})_b) \end{pmatrix}. \tag{50}$$

In this case, the special orthogonal transformations $(S_I)_i{}^m$ leave motionless the identity vector $e = \frac{1}{\sqrt{2}}\eta_0$. Hence, the analogical transformations $(S_I)_a{}^b$ leave motionless the vector $i = ei = \frac{1}{\sqrt{2}}\eta_{n/2}$. Thus, the special orthogonal transformations have the form

$$(S_I)_\alpha{}^\beta := \begin{pmatrix} (S_I)_m{}^i & 0 \\ 0 & (S_I)_c{}^a \end{pmatrix}. \tag{51}$$

2.

$$\begin{pmatrix} 0 & -(h_I)_{ia} + \frac{1}{2}(\eta_I)_i(\eta_{n/2})_a \\ -(h_I)_{bj} - \frac{1}{2}(\eta_{n/2})_b(\eta_I)_j & 0 \end{pmatrix}. \tag{52}$$

In this case, the special orthogonal transformations leave motionless the identity vector and the vector rail line of $i$ with changing the direction, convert the vector $i_1$ to the vector $i_{I+n/2}$ and vice versa, and have the form $(S_{I+\frac{n}{2}})_\alpha{}^\beta := (\tilde{S}_I)_\alpha{}^\gamma (S_I)_\gamma{}^\beta$, where

$$(\tilde{S}_I)_\alpha{}^\beta := \begin{pmatrix} -\frac{1}{\sqrt{2}}\delta_i{}^k + \frac{1}{2\sqrt{2}}(\eta_I)_i(\eta_I)^k + \frac{1}{2}(1+\frac{1}{\sqrt{2}})(\eta_0)_i(\eta_0)^k & \frac{1}{\sqrt{2}}\delta_i{}^a + \frac{1}{2}(1-\frac{1}{\sqrt{2}})(\eta_I)_i(\eta_{I+n/2})^a - \frac{1}{2\sqrt{2}}(\eta_0)_i(\eta_{n/2})^a \\ \frac{1}{\sqrt{2}}\delta_b{}^k + \frac{1}{2}(1-\frac{1}{\sqrt{2}})(\eta_{I+n/2})_b(\eta_I)^k - \frac{1}{2\sqrt{2}}(\eta_{n/2})_b(\eta_0)^k & \frac{1}{\sqrt{2}}\delta_b{}^a - \frac{1}{2\sqrt{2}}(\eta_{I+n/2})_b(\eta_{I+n/2})^a - \frac{1}{2}(1+\frac{1}{\sqrt{2}})(\eta_{n/2})_b(\eta_{n/2})^a \end{pmatrix}. \tag{53}$$

3.
In this case, the special orthogonal transformation $S_{n/2}$ leave motionless all the vectors with $\begin{cases} \text{even index r}, & r < n/2; \\ \text{odd index r}, & r \geq n/2. \end{cases}$ Then the remaining transformations are such that $(< i_1, i_1 >= -1)$:

$$\begin{cases} -i_1 & \to & i_{n/2}, & r=1; \\ i_{n/2} & \to & i_1, & r = n/2; \\ < i_r i_{r-1}, i_1 > i_r & \to & i_{n/2+r-1}, & r = 2s+1 < n/2,\ r>1; \\ < i_{r+1} i_r, i_1 > i_r & \to & i_{r-n/2+1}, & r = 2s > n/2. \end{cases} \tag{54}$$

□

Thus, all is made necessary for a technical realization of Algorithm 1.

**Example 2.** *Let n=16. Algorithm 1 is realized in Appendix.*

1. *This article contains the file "sedenion.pas" (by the operator \ input{sedenion.pas}) being a programming unit adapted to the LaTex (LaTex version of this article is on arXiv:1204.0194) for the Delphi. At the same time, this file is Appendix to this article. You must create a project with this "unit sedenion" and put on the form Button1: TButton, StringGrid1: TStringGrid (the lines 22-24).*

2. *At error occurrence "Stack overflow", it is necessary to adjust the line 15.*

3. *At the lines 146-163, the connecting operators* $\sqrt{2}\eta_i{}^{AB}$ *for* $\mathbb{R}^8$ $(i, A, B = \overline{1,8})$ *are constructed.*

4. *At the lines 164-166, the metric spinor* $\varepsilon_{AB}$ *is constructed.*

5. *At the lines 167-183, the connecting operators* $\sqrt{2}\eta_i{}^A{}_B$, $\sqrt{2}\eta_{iA}{}^B$, $\sqrt{2}(\eta^T)_i{}^A{}_B$, $\sqrt{2}(\eta^T)_{iA}{}^B$ *are constructed.*

6. *At the lines 186-251, the connecting operators, multiplied by* $\sqrt{2}$, *are constructed according to the step 1 of Algorithm 1 for n=14.*

7. *At the lines 252-289, the connecting operators are constructed according to the step 2 of Algorithm 1 for n=16.*

8. *At the lines 290-294, the controlling spinor* $X^A$ *is constructed according to (34).*

9. *At the lines 295-303, the inclusion operator* $P_j{}^A := \eta_{jAB} X^A$ *is constructed according to (38).*

10. *At the lines 304-333, the structural constants* $(\eta_{gen})_{ij}{}^k := \sqrt{2}\eta_i{}^{AB}\ P_{iA} P^k{}_B$ *are constructed according to (5),(35).*

11. *At the lines 334-479, the basic orthogonal transformations* $S_I$ *are constructed according to (29).*

12. *At the lines 480-534, the canonical sedenion structural constants are constructed according to (29).*

13. *At the lines 535-548, the canonical sedenion structural constants are outputted.*

14. *At the lines 549-584, the canonical sedenion structural constants are outputted into the file:*

Table 3: The canonical sedenion multiplication table.

| * | $e_0$ | $e_1$ | $e_2$ | $e_3$ | $e_4$ | $e_5$ | $e_6$ | $e_7$ | $e_8$ | $e_9$ | $e_{10}$ | $e_{11}$ | $e_{12}$ | $e_{13}$ | $e_{14}$ | $e_{15}$ |
|---|---|---|---|---|---|---|---|---|---|---|---|---|---|---|---|---|
| $e_0$ | $e_0$ | $e_1$ | $e_2$ | $e_3$ | $e_4$ | $e_5$ | $e_6$ | $e_7$ | $e_8$ | $e_9$ | $e_{10}$ | $e_{11}$ | $e_{12}$ | $e_{13}$ | $e_{14}$ | $e_{15}$ |
| $e_1$ | $e_1$ | $-e_0$ | $e_3$ | $-e_2$ | $e_5$ | $-e_4$ | $-e_7$ | $e_6$ | $e_9$ | $-e_8$ | $-e_{11}$ | $e_{10}$ | $-e_{13}$ | $e_{12}$ | $e_{15}$ | $-e_{14}$ |
| $e_2$ | $e_2$ | $-e_3$ | $-e_0$ | $e_1$ | $e_6$ | $e_7$ | $-e_4$ | $-e_5$ | $e_{10}$ | $e_{11}$ | $-e_8$ | $-e_9$ | $-e_{14}$ | $-e_{15}$ | $e_{12}$ | $e_{13}$ |
| $e_3$ | $e_3$ | $e_2$ | $-e_1$ | $-e_0$ | $e_7$ | $-e_6$ | $e_5$ | $-e_4$ | $e_{11}$ | $-e_{10}$ | $e_9$ | $-e_8$ | $-e_{15}$ | $e_{14}$ | $-e_{13}$ | $e_{12}$ |
| $e_4$ | $e_4$ | $-e_5$ | $-e_6$ | $-e_7$ | $-e_0$ | $e_1$ | $e_2$ | $e_3$ | $e_{12}$ | $e_{13}$ | $e_{14}$ | $e_{15}$ | $-e_8$ | $-e_9$ | $-e_{10}$ | $-e_{11}$ |
| $e_5$ | $e_5$ | $e_4$ | $-e_7$ | $e_6$ | $-e_1$ | $-e_0$ | $-e_3$ | $e_2$ | $e_{13}$ | $-e_{12}$ | $e_{15}$ | $-e_{14}$ | $e_9$ | $-e_8$ | $e_{11}$ | $-e_{10}$ |
| $e_6$ | $e_6$ | $e_7$ | $e_4$ | $-e_5$ | $-e_2$ | $e_3$ | $-e_0$ | $-e_1$ | $e_{14}$ | $-e_{15}$ | $-e_{12}$ | $e_{13}$ | $e_{10}$ | $-e_{11}$ | $-e_8$ | $e_9$ |
| $e_7$ | $e_7$ | $-e_6$ | $e_5$ | $e_4$ | $-e_3$ | $-e_2$ | $e_1$ | $-e_0$ | $e_{15}$ | $e_{14}$ | $-e_{13}$ | $-e_{12}$ | $e_{11}$ | $e_{10}$ | $-e_9$ | $-e_8$ |
| $e_8$ | $e_8$ | $-e_9$ | $-e_{10}$ | $-e_{11}$ | $-e_{12}$ | $-e_{13}$ | $-e_{14}$ | $-e_{15}$ | $-e_0$ | $e_1$ | $e_2$ | $e_3$ | $e_4$ | $e_5$ | $e_6$ | $e_7$ |
| $e_9$ | $e_9$ | $e_8$ | $-e_{11}$ | $e_{10}$ | $-e_{13}$ | $e_{12}$ | $e_{15}$ | $-e_{14}$ | $-e_1$ | $-e_0$ | $-e_3$ | $e_2$ | $-e_5$ | $e_4$ | $e_7$ | $-e_6$ |
| $e_{10}$ | $e_{10}$ | $e_{11}$ | $e_8$ | $-e_9$ | $-e_{14}$ | $-e_{15}$ | $e_{12}$ | $e_{13}$ | $-e_2$ | $e_3$ | $-e_0$ | $-e_1$ | $-e_6$ | $-e_7$ | $e_4$ | $e_5$ |
| $e_{11}$ | $e_{11}$ | $-e_{10}$ | $e_9$ | $e_8$ | $-e_{15}$ | $e_{14}$ | $-e_{13}$ | $e_{12}$ | $-e_3$ | $-e_2$ | $e_1$ | $-e_0$ | $-e_7$ | $e_6$ | $-e_5$ | $e_4$ |
| $e_{12}$ | $e_{12}$ | $e_{13}$ | $e_{14}$ | $e_{15}$ | $e_8$ | $-e_9$ | $-e_{10}$ | $-e_{11}$ | $-e_4$ | $e_5$ | $e_6$ | $e_7$ | $-e_0$ | $-e_1$ | $-e_2$ | $-e_3$ |
| $e_{13}$ | $e_{13}$ | $-e_{12}$ | $e_{15}$ | $-e_{14}$ | $e_9$ | $e_8$ | $e_{11}$ | $-e_{10}$ | $-e_5$ | $-e_4$ | $e_7$ | $-e_6$ | $e_1$ | $-e_0$ | $e_3$ | $-e_2$ |
| $e_{14}$ | $e_{14}$ | $-e_{15}$ | $-e_{12}$ | $e_{13}$ | $e_{10}$ | $-e_{11}$ | $e_8$ | $e_9$ | $-e_6$ | $-e_7$ | $-e_4$ | $e_5$ | $e_2$ | $-e_3$ | $-e_0$ | $e_1$ |
| $e_{15}$ | $e_{15}$ | $e_{14}$ | $-e_{13}$ | $-e_{12}$ | $e_{11}$ | $e_{10}$ | $-e_9$ | $e_8$ | $-e_7$ | $e_6$ | $-e_5$ | $-e_4$ | $e_3$ | $e_2$ | $-e_1$ | $-e_0$ |

15. *Algorithm 1 is realized by the scheme "one-to-one" and therefore, it requires more memory allocation (the line 15). In order to apply it to higher dimensions, it should be optimized.*

16. *System characteristics of the computer on which the program is tested: HP Pavilion dv7-6b50er i3-2330M/4096/500/Radeon HD6770 2Gb/Win7 HP64*

17. *Run time <3 sec.*

**Note 4.** *The orthogonal transformations* $S_i^{\ j}$ *from the group* $O(n,\mathbb{R})$ $(SO(n,\mathbb{R}))$ *generate the pinor (spinor) transformations* $S_A^{\ B}$ *from the group* $pin(n,\mathbb{R})$ $(Spin(n,\mathbb{R}))$ *which are allocated with the real structure by the involution* $S_A^{\ B'}$ *according to (6.47) from [3, p. 29 (eng), p. 156(33) (rus)]. The pinor (spinor) transformations represent the subgroup of the orthogonal group* $O_{\mathbb{R}}(2^{\frac{n}{2}-1},\mathbb{C})$ $(SO_{\mathbb{R}}(2^{\frac{n}{2}-1},\mathbb{C}))$ *(in the sense* $S_A^{\ B}S_C^{\ D}\varepsilon_{BD}=\varepsilon_{AC}$, $\bar{S}_{A'}^{\ B'}=\bar{S}_{A'}^{\ C}S_C^{\ D}S_D^{\ B'}$, *where* $\varepsilon_{BD}$ *is the metric spinor). The orthogonal transformations from the group* $O(n,\mathbb{R})$ $(SO(n,\mathbb{R}))$ *keeping the algebra identity cause the transformations of the controlling spinor* $\theta^{CD}$ *without changing the connecting operators* $\eta_i^{\ AB}$*. Therefore, the quotient group* $O_{\mathbb{R}}(2^{\frac{n}{2}-1},\mathbb{C})/pin(n,\mathbb{R})$ $(SO_{\mathbb{R}}(2^{\frac{n}{2}-1},\mathbb{C})/Spin(n,\mathbb{R}))$ *will implement the classification of such the hypercomplex orthogonal homogenous algebras* $\mathbb{A}$*. Besides, the classification is carried out on eigenvalues of the controlling spin-tensor* $\theta^{CD}$ *(48) because any symmetric spinor* $\theta^{CD}$ *(48) is led to a diagonal form by the orthogonal transformation from the group* $SO_{\mathbb{R}}(2^{\frac{n}{2}-1},\mathbb{C})$*.*

Let's go back to the axioms. In fact, Axiom 5 is redundant.

*Proof.* From Axiom 4 $(aa)b-a(ab)=b(aa)-(ba)a$, replacing $a$ on $x+y$ according to [6, l. 15, p. 322], we will obtain

$$\begin{gathered}(xy)b+(yx)b-x(yb)-y(xb)+\\ +\underbrace{(xx)b+(yy)b-x(xb)-y(yb)=b(xx)+b(yy)-(bx)x-(by)y}_{\text{Axiom 4}}+\\ +b(xy)+b(yx)-(bx)y-(by)x.\end{gathered}\tag{55}$$

Applying Axiom 4 and setting $y=b$, we will come to

$$\begin{gathered}(bx)b-b(xb)+\underbrace{(xb)b-x(bb)=b(bx)-(bb)x}_{\text{Axiom 4}}+b(xb)-(bx)b,\\ (bx)b-b(xb)=b(xb)-(bx)b,\\ (bx)b=b(xb).\end{gathered}\tag{56}$$

Replacing $b:=y+z$, one can obtain

$$((y+z)x)(y+z)=(y+z)(x(y+z)),\quad (zx)y+(yx)z=z(xy)+y(xz).\tag{57}$$

□

However, if we have the power associativity ($aa^2=a^2a,\quad a:=x+y$) and the flexible identity (Axiom 5) then Axiom 4 is redundant.

From Axiom 4, we can obtain the following identities:

1. Applying Axiom 4, one can obtain

$$\begin{gathered}(y^2x-xy^2)y=(y(yx)-(xy)y)y=y(y(xy))-((xy)y)y=y^2(xy)-(xy)y^2,\\ y(y^2x-xy^2)=y(y(yx)-(xy)y)=y(y(yx))-((yx)y)y=(yx)y^2-y^2(yx).\end{gathered}\tag{58}$$

2. Applying (57) with $z:=a,\quad x:=a^2$, one can obtain

$$(ya^2)a-a(a^2y)=ya^3-a^3y.\tag{59}$$

3. Applying (57) with $x := a, \quad z := a^2$, one can obtain

$$(ya)a^2 - a^2(ay) = ya^3 - a^3y. \tag{60}$$

4. Applying (57), one can obtain

(a)

$$\begin{cases} ((ab)a)x + (xa)(ab) = (ab)(ax) + x(a(ab)), \\ -a(b(ax)) - (ax)(ba) = -(ab)(ax) - ((ax)b)a, \end{cases}$$
$$((ab)a)x - a(b(ax)) = x(a(ab)) - ((ax)b)a + (ax)(ba) - (xa)(ab). \tag{61}$$

(b)

$$\begin{cases} x(a(ba)) + (ba)(ax) = (xa)(ba) + ((ba)a)x, \\ -((xa)b)a - (ab)(xa) = -(xa)(ba) - a(b(xa)), \end{cases}$$
$$x(a(ba)) - ((xa)b)a = ((ba)a)x - a(b(xa)) + (ab)(xa) - (ba)(ax). \tag{62}$$

(c)

$$\begin{array}{c} ((ab)a)x - a(b(ax)) - x(a(ba)) + ((xa)b)a = \\ = x(a(ab)) - ((ax)b)a + (ax)(ba) - (xa)(ab) - \\ -((ba)a)x + a(b(xa)) - (ab)(xa) + (ba)(ax), \end{array} \tag{63}$$

$$\begin{array}{c} ((a,b)a)x - a(b(a,x)) - x(a(b,a)) + ((x,a)b)a = \\ = (a,x)(ba) - (xa)(a,b) - (ab)(x,a) + (b,a)(ax). \end{array}$$

Let us remind that $(a,b) := \frac{1}{2}(ab+ba)$ (2), and if $(a,b) = -< a,b > e+$ $+ < a,e > b+ < b,e > a$ then the identity (63) is performed automatically because

$$\begin{array}{c} ((a,b)a)x + (xa)(a,b) = (a,b)(ax) + x(a(a,b)), \\ a(b(a,x)) + (a,x)(ba) = (ab)(a,x) + ((a,x)b)a \end{array} \tag{64}$$

on the base (57). In the equation (61), one can see the Moufang identity (left and right) for n=8 because in this case,

$$x(a(a,b)) - ((a,x)b)a + (a,x)(ba) - (xa)(b,a) = 0, \tag{65}$$

and then

$$((ab)a)x - a(b(ax)) = -x(a(ba)) + ((xa)b)a. \tag{66}$$

Note, the last identities (65)-(66) is executed for n>8 under the condition $< a,e > = < b,e > = 0$.

# Appendix.

//The sedenion multiplication table.

```
{****************************************************************************************}
{                                                                                        }
{ Sedenion Unit for Delphi                                                               }
{                                                                                        }
{ Copyright (c) 2011 Konstantin Andreev All Rights Reserved                              }
{ THIS SOFTWARE IS PROVIDED "AS IS" WITHOUT WARRANTY OF ANY KIND, EITHER EXPRESSED OR IMPLIED, }
{ INCLUDING BUT NOT LIMITED TO THE IMPLIED MERCHANTABILITY AND/OR FITNESS FOR A PARTICULAR }
{ PURPOSE. KONSTANTIN ANDREEV CANNOT BE HELD RESPONSIBLE FOR ANY LOSSES, EITHER DIRECT OR }
{ INDIRECT, OF ANY PARTY MAKING USE OF THIS SOFTWARE. IN MAKING USE OF THIS SOFTWARE, YOU }
{ AGREE TO BE BOUND BY THE TERMS AND CONDITIONS FOUND IN THE ACCOMPANYING LICENSE.       }
{                                                                                        }
{****************************************************************************************}
unit sedenion;
{$M 16384,9000000}

interface
uses
  Windows, Messages, SysUtils, Classes, Graphics, Controls, Forms, Dialogs,
  StdCtrls, Grids, IniFiles;
type
  TForm1 = class(TForm)
    Button1: TButton;
    StringGrid1: TStringGrid;
    procedure Button1Click(Sender: TObject);
  end;
var
  Form1: TForm1;
implementation
{$R *.DFM} //

procedure TForm1.Button1Click(Sender: TObject);
type
   complex = record  x,y   : real; end;
var
    eta_16          : array [1..16,1..128,1..128]  of complex;     //The connection operators
                                                                   //for n=16 $\eta_i{}^{AB}$.
    eta_16_         : array [1..16,1..128,1..128]  of complex;     //The connection operators
                                                                   //for n=16 $\eta_{iAB}$.
    eta_8           : array [1..8,1..8,1..8]  of complex;          //The connection operators
                                                                   //for n=8 $\eta_i{}^{AB}$.
    eta_8_          : array [1..8,1..8,1..8]  of complex;          //The connection operators
                                                                   //contracted with the metric
                                                                   //spinor for n=8 $\eta_i{}^A{}_B$.
    eta_8__         : array [1..8,1..8,1..8]  of complex;          //The connection operators
                                                                   //contracted with the metric
                                                                   //spinor for n=8 $\eta_{iA}{}^B$.
    eta_8_T         : array [1..8,1..8,1..8]  of complex;          //The connection operators
                                                                   //contracted with the metric
                                                                   //spinor for n=8 $(\eta^T)_i{}^A{}_B$.
    eta_8__T        : array [1..8,1..8,1..8]  of complex;          //The connection operators
                                                                   //contracted with the metric
                                                                   //spinor for n=8 $(\eta^T)_{iA}{}^B$.
    eta             : array [1..16,1..16,1..16]  of complex;       //The structural constant of
                                                                   //a generating hypercomplex
                                                                   //algebra for n=16 $(\eta_{gen})_{ij}{}^k$ from (41).
    eta_            : array [1..16,1..128,1..16]  of complex;      //The auxiliary variable
                                                                   //for n=16 $\eta_i{}^{Ak}$.
    e_8             : array [1..8,1..8]  of complex;               //The metric spinor for n=16
                                                                   //$\| \varepsilon_{AB} \| = \| \varepsilon^{AB} \|$.
    P               : array [1..16,1..128]  of complex;            //The inclusion operator
                                                                   //for n=16
                                                                   //$\| P_{iA} \| := \| P^i{}_A \| = \| \eta_{iBA} X^B \|$.
    x               : array [1..128]  of complex;                  //The controlling spinor
                                                                   //for n=16 $X^A$.
    unit_           : array [0..15]  of complex;                   //The  algebra identity $e = \frac{1}{\sqrt{2}}\eta^i$,
                                                                   //$\| \eta^i \| = \| \eta_i \|$.
    g               : array [0..15,0..15]  of complex;             //The metric of the Euclidean
                                                                   //space $\mathbb{R}^{16}$
                                                                   //$\| g_{ij} \| = \| g^{ij} \| = \| \delta_i{}^j \|$.
    i,j,k,l,m,r     : integer;                                     //Indices of an array element.
    mi              : complex;                                     //The complex factor $i$.
    m_i             : complex;                                     //The complex factor $-i$.
    m_              : complex;                                     //The complex factor $-1$.
    m_12            : complex;                                     //The complex factor $-12$.
    m_3             : complex;                                     //The complex factor $\frac{1}{3}$.
    IniFile         : TIniFile;                                    //The output file.
    array_str       : string;                                      //A line of the output file.
    s_orthogonal    : array [1..15,0..15,1..16]  of complex;       //The basic orthogonal
                                                                   //transformations
                                                                   //for n=16 $(S_I)_i{}^j$ $(I = \overline{1,15})$ from note 2.
    eta_orthogonal  : array [1..15,0..16,0..16,0..16]  of complex; //The structural constant
                                                                   //of hypercomplex basic
                                                                   //algebras for n=16 $(\eta_I)_{ij}{}^k$.
    eta_orthogonal_ : array [1..15,0..16,0..16,0..16]  of complex; //The auxiliary variable.
    eta_constant    : array [0..15,0..15,0..15]  of complex;       //The structural constant
                                                                   //of the canonical sedenion
                                                                   //algebra $\eta_{ij}{}^k$.
```

```
 //Addition of complex numbers.
 function add(c11,c12: complex):complex;
 var c13  : complex;
 begin
   c13.x:=c11.x+c12.x;
   c13.y:=c11.y+c12.y;
   add:=c13;
 end;
 //Multiplication of complex numbers.
 function mul(c21,c22: complex):complex;
 var c13  : complex;
 begin
   c13.x:=c21.x*c22.x-c21.y*c22.y;
   c13.y:=c21.x*c22.y+c21.y*c22.x;
   mul:=c13;
 end;
 //Initialization of a complex number.
 procedure Init_(var c31 : complex);
 begin
   c31.x:=0; c31.y:=0;
 end;
begin
//Initialization the connection operators for n=8.
    for i:=1 to 8 do
     for j:=1 to 8 do
      for k:=1 to 8 do
      begin
       Init_(eta_8[i,j,k]);
       Init_(eta_8_[i,j,k]);
      end;
//Initialization the metric spinor for n=8.
    for i:=1 to 8 do
     for j:=1 to 8 do
       Init_(e_8[i,j]);
//Initialization the connection operators for n = 16.
    for i:=1 to 16 do
     for j:=1 to 128 do
      for k:=1 to 128 do
      begin
       Init_(eta_16[i,j,k]);
       Init_(eta_16_[i,j,k]);
      end;
//Initialization the ortogonal transformation for n = 16.
    for m:=1 to 15 do
     for i:=1 to 16 do
      for j:=1 to 16 do
       Init_(s_orthogonal[m,i,j]);
//Initialization the sedenion algebra identity for n=16.
    for i:=0 to 15 do
       Init_(unit_[i]);
       unit_[0].x:=1;
//Initialization the metric tensor.
    for i:=0 to 15 do
    begin
     for j:=0 to 15 do
      Init_(g[i,j]);
     g[i,i].x:=1;
    end;
//Construction the connection operators (generating the octonion algebra) for n=8 (multiplied by √2).
    eta_8[1,1,2].y:=+1; eta_8[1,3,4].y:=+1; eta_8[1,5,6].y:=-1; eta_8[1,7,8].y:=-1;
    eta_8[2,1,2].x:=-1; eta_8[2,3,4].x:=+1; eta_8[2,5,6].x:=-1; eta_8[2,7,8].x:=+1;
    eta_8[3,1,3].y:=-1; eta_8[3,2,4].y:=+1; eta_8[3,5,7].y:=+1; eta_8[3,6,8].y:=-1;
    eta_8[4,1,3].x:=+1; eta_8[4,2,4].x:=+1; eta_8[4,5,7].x:=+1; eta_8[4,6,8].x:=+1;
    eta_8[5,1,4].y:=+1; eta_8[5,2,3].y:=+1; eta_8[5,5,8].y:=-1; eta_8[5,6,7].y:=-1;
    eta_8[6,1,4].x:=-1; eta_8[6,2,3].x:=+1; eta_8[6,5,8].x:=-1; eta_8[6,6,7].x:=+1;
    for i:=1 to 8 do
     for j:=1 to 8 do
      for k:=1 to j do
      begin
       eta_8[i,j,k].x:=-eta_8[i,k,j].x;
       eta_8[i,j,k].y:=-eta_8[i,k,j].y;
      end;
    eta_8[7,1,5].y:=+1; eta_8[7,2,6].y:=+1; eta_8[7,3,7].y:=+1; eta_8[7,4,8].y:=+1;
    eta_8[7,5,1].y:=-1; eta_8[7,6,2].y:=-1; eta_8[7,7,3].y:=-1; eta_8[7,8,4].y:=-1;
    eta_8[8,1,5].x:=+1; eta_8[8,2,6].x:=+1; eta_8[8,3,7].x:=+1; eta_8[8,4,8].x:=+1;
    eta_8[8,5,1].x:=+1; eta_8[8,6,2].x:=+1; eta_8[8,7,3].x:=+1; eta_8[8,8,4].x:=+1;
//Construction the metric spinor for n = 8.
    e_8[1,5].x:=1;e_8[2,6].x:=1;e_8[3,7].x:=1;e_8[4,8].x:=1;
    e_8[5,1].x:=1;e_8[6,2].x:=1;e_8[7,3].x:=1;e_8[8,4].x:=1;
//Construction the connection operators contracted with the metric spinor for n=8 (multiplied by √2).
    for i:=1 to 8 do
     for j:=1 to 8 do
      for k:=1 to 8 do
      begin
       Init_(eta_8_[i,j,k]);
       Init_(eta_8__[i,j,k]);
       Init_(eta_8_T[i,j,k]);
       Init_(eta_8__T[i,j,k]);
       for m:=1 to 8 do
       begin
         eta_8_[i,j,k]:=add(eta_8_[i,j,k],mul(eta_8[i,j,m],e_8[m,k]));
         eta_8_T[i,j,k]:=add(eta_8_T[i,j,k],mul(eta_8[i,m,j],e_8[m,k]));
         eta_8__[i,j,k]:=add(eta_8__[i,j,k],mul(eta_8[i,m,k],e_8[m,j]));
         eta_8__T[i,j,k]:=add(eta_8__T[i,j,k],mul(eta_8[i,k,m],e_8[m,j]));
       end;
      end;
```

```
//The constant factors: i(mi), -i(m_i), -1(m_).
    mi.x:=0;  mi.y:=1; m_i.x:=0; m_i.y:=-1; m_.x:=-1; m_.y:=0;
//Construction the connection operators for n=14 from (30),(30') (multiplied by √2).
  //1. n=13,14.
    for i:=1 to 8 do
     for l:=1 to 8 do
     begin
       eta_16[13,i+ 0,l+24]:=e_8[i,l];               eta_16[14,i+ 0,l+24]:=mul(mi,e_8[i,l]);
       eta_16[13,i+ 8,l+16]:=mul(m_,e_8[i,l]);       eta_16[14,i+ 8,l+16]:=mul(m_i,e_8[i,l]);

       eta_16[13,i+32,l+56]:=e_8[i,l];               eta_16[14,i+32,l+56]:=mul(m_i,e_8[i,l]);
       eta_16[13,i+40,l+48]:=mul(m_,e_8[i,l]);       eta_16[14,i+40,l+48]:=mul(mi,e_8[i,l]);

       eta_16_[13,i+ 0,l+24]:=e_8[i,l];              eta_16_[14,i+ 0,l+24]:=mul(m_i,e_8[i,l]);
       eta_16_[13,i+ 8,l+16]:=mul(m_,e_8[i,l]);      eta_16_[14,i+ 8,l+16]:=mul(mi,e_8[i,l]);

       eta_16_[13,i+32,l+56]:=e_8[i,l];              eta_16_[14,i+32,l+56]:=mul(mi,e_8[i,l]);
       eta_16_[13,i+40,l+48]:=mul(m_,e_8[i,l]);      eta_16_[14,i+40,l+48]:=mul(m_i,e_8[i,l]);
     end;
  //2. n=11,12.
    for i:=1 to 8 do
     for l:=1 to 8 do
     begin
       eta_16[11,i+ 0,l+40]:=e_8[i,l];                eta_16[12,i+ 0,l+40]:=mul(mi,e_8[i,l]);
       eta_16[11,i+ 8,l+32]:=mul(m_,e_8[i,l]);        eta_16[12,i+ 8,l+32]:=mul(m_i,e_8[i,l]);

       eta_16[11,i+16,l+56]:=mul(m_,e_8[i,l]);        eta_16[12,i+16,l+56]:=mul(mi,e_8[i,l]);
       eta_16[11,i+24,l+48]:=e_8[i,l];                eta_16[12,i+24,l+48]:=mul(m_i,e_8[i,l]);

       eta_16_[11,i+ 0,l+40]:=e_8[i,l];               eta_16_[12,i+ 0,l+40]:=mul(m_i,e_8[i,l]);
       eta_16_[11,i+ 8,l+32]:=mul(m_,e_8[i,l]);       eta_16_[12,i+ 8,l+32]:=mul(mi,e_8[i,l]);

       eta_16_[11,i+16,l+56]:=mul(m_,e_8[i,l]);       eta_16_[12,i+16,l+56]:=mul(m_i,e_8[i,l]);
       eta_16_[11,i+24,l+48]:=e_8[i,l];               eta_16_[12,i+24,l+48]:=mul(mi,e_8[i,l]);
     end;
  //3. n=9,10.
    for i:=1 to 8 do
     for l:=1 to 8 do
     begin
       eta_16[9,i+ 0,l+48]:=mul(m_i,e_8[i,l]);     eta_16[10,i+ 0,l+48]:=mul(m_,e_8[i,l]);
       eta_16[9,i+ 8,l+56]:=mul(m_i,e_8[i,l]);     eta_16[10,i+ 8,l+56]:=e_8[i,l];

       eta_16[9,i+16,l+32]:=mul(m_i,e_8[i,l]);      eta_16[10,i+16,l+32]:=mul(m_,e_8[i,l]);
       eta_16[9,i+24,l+40]:=mul(m_i,e_8[i,l]);      eta_16[10,i+24,l+40]:=e_8[i,l];

       eta_16_[9,i+ 0,l+48]:=mul(mi,e_8[i,l]);     eta_16_[10,i+ 0,l+48]:=mul(m_,e_8[i,l]);
       eta_16_[9,i+ 8,l+56]:=mul(mi,e_8[i,l]);     eta_16_[10,i+ 8,l+56]:=e_8[i,l];

       eta_16_[9,i+16,l+32]:=mul(mi,e_8[i,l]);     eta_16_[10,i+16,l+32]:=mul(m_,e_8[i,l]);
       eta_16_[9,i+24,l+40]:=mul(mi,e_8[i,l]);     eta_16_[10,i+24,l+40]:=e_8[i,l];
     end;
  //4. n=1-8.
    for i:=1 to 8 do
     for l:=1 to 8 do
      for k:=1 to 8 do
      begin
        eta_16[k,i+ 0,l+56]:=eta_8_[k,i,l];
        eta_16[k,i+ 8,l+48]:=eta_8__T[k,i,l];

        eta_16[k,i+16,l+40]:=eta_8__T[k,i,l];
        eta_16[k,i+24,l+32]:=eta_8_[k,i,l];

        eta_16_[k,i+ 0,l+56]:=eta_8__[k,i,l];
        eta_16_[k,i+ 8,l+48]:=eta_8_T[k,i,l];

        eta_16_[k,i+16,l+40]:=eta_8_T[k,i,l];
        eta_16_[k,i+24,l+32]:=eta_8__[k,i,l];
      end;
//Construction the connection operators for n=16 from (32).
  //1. The skew-symmetry.
    for i:=1 to 16 do
     for j:=1 to 128 do
      for k:=1 to j do
      begin
        eta_16[i,j,k].x:=-eta_16[i,k,j].x;
        eta_16[i,j,k].y:=-eta_16[i,k,j].y;
        eta_16_[i,j,k].x:=-eta_16_[i,k,j].x;
        eta_16_[i,j,k].y:=-eta_16_[i,k,j].y;
      end;
  //2. Duplication.
    for i:=1 to 64 do
     for l:=1 to 64 do
      for k:=1 to 14 do
      begin
        eta_16[k,i+64,l+64]:=mul(m_,eta_16_[k,l,i]);
        eta_16_[k,i+64,l+64]:=mul(m_,eta_16[k,l,i]);
      end;
  //3. n=15,16.
    for i:=1 to 64 do
    begin
      eta_16[15,i,i+64].x:=0;   eta_16[15,i,i+64].y:=1;    eta_16[16,i,i+64].x:=1;    eta_16[16,i,i+64].y:=0;
      eta_16[15,i+64,i].x:=0;   eta_16[15,i+64,i].y:=-1;   eta_16[16,i+64,i].x:=1;    eta_16[16,i+64,i].y:=0;
      eta_16_[15,i,i+64].x:=0;  eta_16_[15,i,i+64].y:=-1;  eta_16_[16,i,i+64].x:=1;   eta_16_[16,i,i+64].y:=0;
      eta_16_[15,i+64,i].x:=0;  eta_16_[15,i+64,i].y:=1;   eta_16_[16,i+64,i].x:=1;   eta_16_[16,i+64,i].y:=0;
    end;

```

```
  //4. Multiply by 1/√2.
    for i:=1 to 16 do
     for j:=1 to 128 do
      for k:=1 to 128 do
      begin
        eta_16[i,j,k].x:=eta_16[i,j,k].x/sqrt(2);
        eta_16[i,j,k].y:=eta_16[i,j,k].y/sqrt(2);
        eta_16_[i,j,k].x:=eta_16_[i,j,k].x/sqrt(2);
        eta_16_[i,j,k].y:=eta_16_[i,j,k].y/sqrt(2);
      end;
//Initialization the controlling spinors X^A from (34).
    for j:=1 to 128 do
      Init_(x[j]);
    x[1].x:=1;
    x[65].x:=1;
//Construction of the inclusion operators P_jA from (38).
    for l:=1 to 4 do
     for i:=1 to 16 do
      for j:=1 to 128 do
      begin
        Init_(P[i,j]);
        for k:=1 to 128 do
          P[i,j]:=add(P[i,j],mul(eta_16_[i,k,j],x[k]));
      end;
//Construction of the structure constants of the generating algebra for n=16 A_gen((η_gen)_ij^k) from (5),(35).
  //1. Multiply the connection operators by P.
    begin
     for i:=1 to 16 do
      for j:=1 to 128 do
       for k:=1 to 16 do
       begin
         eta_[i,j,k].x:=0;
         eta_[i,j,k].y:=0;
         for m:=1 to 128 do
           eta_[i,j,k]:=add(eta_[i,j,k],mul(eta_16[i,j,m],P[k,m]));
       end;
     for i:=1 to 16 do
      for j:=1 to 16 do
       for k:=1 to 16 do
       begin
         eta[i,j,k].x:=0;
         eta[i,j,k].y:=0;
         for m:=1 to 128 do
           eta[i,j,k]:=add(eta[i,j,k],mul(eta_[i,m,k],P[j,m]));
       end;
    end;
 //2. Multiply by √2.
    for i:=1 to 16 do
     for j:=1 to 16 do
      for k:=1 to 16 do
      begin
         eta[i,j,k].x:=eta[i,j,k].x*sqrt(2);
         eta[i,j,k].y:=eta[i,j,k].y*sqrt(2);
      end;
//Initialization the basic orthogonal transformations (S_I)_i^j (I = 1,15) according to note 2.
 //I=1.
     s_orthogonal[1,3,1].x:=-1;      s_orthogonal[1,2,2].x:=1;
     s_orthogonal[1,5,3].x:=-1;      s_orthogonal[1,4,4].x:=1;
     s_orthogonal[1,7,5].x:=1;       s_orthogonal[1,6,6].x:=1;
     s_orthogonal[1,9,7].x:=1;       s_orthogonal[1,8,8].x:=1;
     s_orthogonal[1,11,9].x:=-1;     s_orthogonal[1,10,10].x:=1;
     s_orthogonal[1,13,11].x:=1;     s_orthogonal[1,12,12].x:=1;
     s_orthogonal[1,15,13].x:=-1;    s_orthogonal[1,14,14].x:=1;
     s_orthogonal[1,1,15].x:=-1;     s_orthogonal[1,0,16].x:=1;
 //I=2.
     s_orthogonal[2,12,1].x:=-1;     s_orthogonal[2,14,2].x:=1;
     s_orthogonal[2,1,3].x:=-1;      s_orthogonal[2,3,4].x:=1;
     s_orthogonal[2,7,5].x:=1;       s_orthogonal[2,5,6].x:=-1;
     s_orthogonal[2,6,7].x:=1;       s_orthogonal[2,4,8].x:=1;
     s_orthogonal[2,11,9].x:=-1;     s_orthogonal[2,9,10].x:=-1;
     s_orthogonal[2,10,11].x:=1;     s_orthogonal[2,8,12].x:=-1;
     s_orthogonal[2,15,13].x:=-1;    s_orthogonal[2,13,14].x:=-1;
     s_orthogonal[2,2,15].x:=-1;     s_orthogonal[2,0,16].x:=1;
 //I=3.
     s_orthogonal[3,1,1].x:=1;       s_orthogonal[3,2,2].x:=1;
     s_orthogonal[3,7,3].x:=-1;      s_orthogonal[3,4,4].x:=1;
     s_orthogonal[3,5,5].x:=-1;      s_orthogonal[3,6,6].x:=1;
     s_orthogonal[3,11,7].x:=1;      s_orthogonal[3,8,8].x:=1;
     s_orthogonal[3,9,9].x:=1;       s_orthogonal[3,10,10].x:=1;
     s_orthogonal[3,15,11].x:=1;     s_orthogonal[3,12,12].x:=1;
     s_orthogonal[3,13,13].x:=1;     s_orthogonal[3,14,14].x:=1;
     s_orthogonal[3,3,15].x:=-1;     s_orthogonal[3,0,16].x:=1;
 //I=4.
     s_orthogonal[4,1,1].x:=1;       s_orthogonal[4,5,2].x:=-1;
     s_orthogonal[4,3,3].x:=1;       s_orthogonal[4,7,4].x:=-1;
     s_orthogonal[4,2,5].x:=-1;      s_orthogonal[4,6,6].x:=1;
     s_orthogonal[4,12,7].x:=1;      s_orthogonal[4,8,8].x:=1;
     s_orthogonal[4,9,9].x:=-1;      s_orthogonal[4,13,10].x:=1;
     s_orthogonal[4,11,11].x:=1;     s_orthogonal[4,15,12].x:=1;
     s_orthogonal[4,10,13].x:=1;     s_orthogonal[4,14,14].x:=1;
     s_orthogonal[4,4,15].x:=-1;     s_orthogonal[4,0,16].x:=1;
```

```
//I=5.
    s_orthogonal[5,3,1].x:=1;      s_orthogonal[5,6,2].x:=1;
    s_orthogonal[5,1,3].x:=1;      s_orthogonal[5,4,4].x:=1;
    s_orthogonal[5,13,5].x:=-1;    s_orthogonal[5,8,6].x:=1;
    s_orthogonal[5,7,7].x:=-1;     s_orthogonal[5,2,8].x:=1;
    s_orthogonal[5,9,9].x:=1;      s_orthogonal[5,12,10].x:=1;
    s_orthogonal[5,15,11].x:=-1;   s_orthogonal[5,10,12].x:=1;
    s_orthogonal[5,11,13].x:=1;    s_orthogonal[5,14,14].x:=-1;
    s_orthogonal[5,5,15].x:=-1;    s_orthogonal[5,0,16].x:=1;
//I=6.
    s_orthogonal[6,8,1].x:=-1;     s_orthogonal[6,14,2].x:=-1;
    s_orthogonal[6,3,3].x:=1;      s_orthogonal[6,5,4].x:=-1;
    s_orthogonal[6,4,5].x:=1;      s_orthogonal[6,2,6].x:=-1;
    s_orthogonal[6,7,7].x:=1;      s_orthogonal[6,1,8].x:=1;
    s_orthogonal[6,9,9].x:=1;      s_orthogonal[6,15,10].x:=1;
    s_orthogonal[6,11,11].x:=1;    s_orthogonal[6,13,12].x:=1;
    s_orthogonal[6,12,13].x:=1;    s_orthogonal[6,10,14].x:=1;
    s_orthogonal[6,6,15].x:=-1;    s_orthogonal[6,0,16].x:=1;
//I=7.
    s_orthogonal[7,1,1].x:=1;      s_orthogonal[7,6,2].x:=-1;
    s_orthogonal[7,3,3].x:=1;      s_orthogonal[7,4,4].x:=1;
    s_orthogonal[7,5,5].x:=-1;     s_orthogonal[7,2,6].x:=1;
    s_orthogonal[7,15,7].x:=1;     s_orthogonal[7,8,8].x:=1;
    s_orthogonal[7,9,9].x:=1;      s_orthogonal[7,14,10].x:=-1;
    s_orthogonal[7,11,11].x:=1;    s_orthogonal[7,12,12].x:=-1;
    s_orthogonal[7,13,13].x:=1;    s_orthogonal[7,10,14].x:=1;
    s_orthogonal[7,7,15].x:=-1;    s_orthogonal[7,0,16].x:=1;
//I=8.
    s_orthogonal[8,1,1].x:=1;      s_orthogonal[8,9,2].x:=-1;
    s_orthogonal[8,3,3].x:=1;      s_orthogonal[8,11,4].x:=-1;
    s_orthogonal[8,5,5].x:=1;      s_orthogonal[8,13,6].x:=-1;
    s_orthogonal[8,7,7].x:=1;      s_orthogonal[8,15,8].x:=1;
    s_orthogonal[8,2,9].x:=1;      s_orthogonal[8,10,10].x:=1;
    s_orthogonal[8,4,11].x:=1;     s_orthogonal[8,12,12].x:=-1;
    s_orthogonal[8,6,13].x:=1;     s_orthogonal[8,14,14].x:=-1;
    s_orthogonal[8,8,15].x:=-1;    s_orthogonal[8,0,16].x:=1;
//I=9.
    s_orthogonal[9,1,1].x:=1;      s_orthogonal[9,8,2].x:=1;
    s_orthogonal[9,3,3].x:=1;      s_orthogonal[9,10,4].x:=1;
    s_orthogonal[9,5,5].x:=1;      s_orthogonal[9,12,6].x:=1;
    s_orthogonal[9,11,7].x:=-1;    s_orthogonal[9,2,8].x:=1;
    s_orthogonal[9,13,9].x:=-1;    s_orthogonal[9,4,10].x:=1;
    s_orthogonal[9,15,11].x:=-1;   s_orthogonal[9,6,12].x:=1;
    s_orthogonal[9,7,13].x:=1;     s_orthogonal[9,14,14].x:=-1;
    s_orthogonal[9,9,15].x:=-1;    s_orthogonal[9,0,16].x:=1;
//I=10.
    s_orthogonal[10,1,1].x:=1;     s_orthogonal[10,11,2].x:=1;
    s_orthogonal[10,3,3].x:=1;     s_orthogonal[10,9,4].x:=-1;
    s_orthogonal[10,5,5].x:=1;     s_orthogonal[10,15,6].x:=-1;
    s_orthogonal[10,7,7].x:=1;     s_orthogonal[10,13,8].x:=-1;
    s_orthogonal[10,4,9].x:=1;     s_orthogonal[10,14,10].x:=1;
    s_orthogonal[10,2,11].x:=1;    s_orthogonal[10,8,12].x:=1;
    s_orthogonal[10,12,13].x:=-1;  s_orthogonal[10,6,14].x:=1;
    s_orthogonal[10,10,15].x:=-1;  s_orthogonal[10,0,16].x:=1;
//I=11.
    s_orthogonal[11,1,1].x:=1;     s_orthogonal[11,10,2].x:=-1;
    s_orthogonal[11,3,3].x:=1;     s_orthogonal[11,8,4].x:=1;
    s_orthogonal[11,5,5].x:=1;     s_orthogonal[11,14,6].x:=1;
    s_orthogonal[11,15,7].x:=-1;   s_orthogonal[11,4,8].x:=1;
    s_orthogonal[11,9,9].x:=1;     s_orthogonal[11,2,10].x:=1;
    s_orthogonal[11,7,11].x:=1;    s_orthogonal[11,12,12].x:=1;
    s_orthogonal[11,13,13].x:=1;   s_orthogonal[11,6,14].x:=1;
    s_orthogonal[11,11,15].x:=-1;  s_orthogonal[11,0,16].x:=1;
//I=12.
    s_orthogonal[12,1,1].x:=1;     s_orthogonal[12,13,2].x:=1;
    s_orthogonal[12,3,3].x:=1;     s_orthogonal[12,15,4].x:=1;
    s_orthogonal[12,5,5].x:=1;     s_orthogonal[12,9,6].x:=-1;
    s_orthogonal[12,7,7].x:=1;     s_orthogonal[12,11,8].x:=1;
    s_orthogonal[12,6,9].x:=1;     s_orthogonal[12,10,10].x:=1;
    s_orthogonal[12,8,11].x:=-1;   s_orthogonal[12,4,12].x:=1;
    s_orthogonal[12,2,13].x:=1;    s_orthogonal[12,14,14].x:=1;
    s_orthogonal[12,12,15].x:=-1;  s_orthogonal[12,0,16].x:=1;
//I=13.
    s_orthogonal[13,1,1].x:=1;     s_orthogonal[13,12,2].x:=-1;
    s_orthogonal[13,3,3].x:=1;     s_orthogonal[13,14,4].x:=-1;
    s_orthogonal[13,5,5].x:=1;     s_orthogonal[13,8,6].x:=1;
    s_orthogonal[13,11,7].x:=1;    s_orthogonal[13,6,8].x:=1;
    s_orthogonal[13,7,9].x:=1;     s_orthogonal[13,10,10].x:=1;
    s_orthogonal[13,15,11].x:=-1;  s_orthogonal[13,2,12].x:=1;
    s_orthogonal[13,9,13].x:=-1;   s_orthogonal[13,4,14].x:=1;
    s_orthogonal[13,13,15].x:=-1;  s_orthogonal[13,0,16].x:=1;
//I=14.
    s_orthogonal[14,1,1].x:=1;     s_orthogonal[14,15,2].x:=-1;
    s_orthogonal[14,3,3].x:=1;     s_orthogonal[14,13,4].x:=1;
    s_orthogonal[14,5,5].x:=1;     s_orthogonal[14,11,6].x:=-1;
    s_orthogonal[14,7,7].x:=1;     s_orthogonal[14,9,8].x:=-1;
    s_orthogonal[14,8,9].x:=1;     s_orthogonal[14,6,10].x:=1;
    s_orthogonal[14,10,11].x:=-1;  s_orthogonal[14,4,12].x:=1;
    s_orthogonal[14,12,13].x:=1;   s_orthogonal[14,2,14].x:=1;
    s_orthogonal[14,14,15].x:=-1;  s_orthogonal[14,0,16].x:=1;
```

```
 //I=15.
     s_orthogonal[15,1,1].x:=1;        s_orthogonal[15,14,2].x:=1;
     s_orthogonal[15,3,3].x:=1;        s_orthogonal[15,12,4].x:=-1;
     s_orthogonal[15,5,5].x:=1;        s_orthogonal[15,10,6].x:=1;
     s_orthogonal[15,7,7].x:=1;        s_orthogonal[15,8,8].x:=-1;
     s_orthogonal[15,9,9].x:=-1;       s_orthogonal[15,6,10].x:=1;
     s_orthogonal[15,11,11].x:=-1;     s_orthogonal[15,4,12].x:=1;
     s_orthogonal[15,13,13].x:=1;      s_orthogonal[15,2,14].x:=1;
     s_orthogonal[15,15,15].x:=-1;     s_orthogonal[15,0,16].x:=1;
//Constructing the structural constant of the canonical sedenion algebra for n=16.
//1. Constructing the structural constants of the basic algebras
//$A_I:(\eta_I)_{ij}{}^k:=(S_I)_i{}^l(S_I)_j{}^m(\eta_{gen})_{lm}{}^r(S_I)^k{}_r$ $(I=\overline{1,15})$ from (29).
    for m:=1 to 15 do
    begin
     for i:=1 to 16 do
      for j:=1 to 16 do
       for k:=0 to 15 do
       begin
         eta_orthogonal[m,i,j,k].x:=0;
         eta_orthogonal[m,i,j,k].y:=0;
         for l:=1 to 16 do
          eta_orthogonal[m,i,j,k]:=
          add(eta_orthogonal[m,i,j,k],mul(s_orthogonal[m,k,l],eta[i,j,l]));
       end;
     for i:=1 to 16 do
      for j:=0 to 15 do
       for k:=0 to 15 do
       begin
         eta_orthogonal_[m,i,j,k].x:=0;
         eta_orthogonal_[m,i,j,k].y:=0;
         for l:=1 to 16 do
          eta_orthogonal_[m,i,j,k]:=
          add(eta_orthogonal_[m,i,j,k],mul(s_orthogonal[m,j,l],eta_orthogonal[m,i,l,k]));
       end;
     for i:=0 to 15 do
      for j:=0 to 15 do
       for k:=0 to 15 do
       begin
         eta_orthogonal[m,i,j,k].x:=0;
         eta_orthogonal[m,i,j,k].y:=0;
         for l:=1 to 16 do
          eta_orthogonal[m,i,j,k]:=
          add(eta_orthogonal[m,i,j,k],mul(s_orthogonal[m,i,l],eta_orthogonal_[m,l,j,k]));
       end;
    end;
//2. Initialization the complex facror $-12(m\_12)$, $\frac{1}{3}(m\_3)$.
    m_12.x:=-12;
    m_12.y:=0;
    m_3.x:=1/3;
    m_3.y:=0;
//3.Constructing according to
//$\eta_{ij}{}^k:=(1-\sum_{I=1}^{15}\frac{1}{3}I)(\eta_0)_{ij}{}^k+\frac{1}{3}\sum_{I=1}^{15}\underbrace{((h_I)_{ij}{}^k+(\eta_0)_{ij}{}^k)}_{(\eta_I)_{ij}{}^k}=\frac{1}{3}(-12(\eta_0)_{ij}{}^k+\sum_{I=1}^{15}(\eta_I)_{ij}{}^k)$,
//$(\eta_0)_{ij}{}^k:=((\frac{1}{\sqrt{2}}\eta_i)\delta_j{}^k+(\frac{1}{\sqrt{2}}\eta_j)\delta_i{}^k-g_{ij}(\frac{1}{\sqrt{2}}\eta^k))$ from (29),(41).
    for i:=0 to 15 do
     for j:=0 to 15 do
      for k:=0 to 15 do
      begin
       eta_constant[i,j,k].x:=0;
       eta_constant[i,j,k].y:=0;
       for m:=1 to 15 do
        eta_constant[i,j,k]:=add(eta_constant[i,j,k],eta_orthogonal[m,i,j,k]);
       eta_constant[i,j,k]:=mul(m_3,add(eta_constant[i,j,k],
              mul(m_12,add(add(mul(unit_[j],g[i,k]),mul(unit_[i],g[j,k])),mul(m_,mul(g[i,j],unit_[k]))))));
      end;
//Output of the sedenion multiplication table.
    for i:=1 to 16 do
    begin
      StringGrid1.Cells[0,i]:=IntToStr(i-1);
      StringGrid1.Cells[i,0]:=IntToStr(i-1);
    end;
    for i:=0 to 15 do
     for j:=0 to 15 do
     begin
       StringGrid1.Cells[j+1,i+1]:='';
       for k:=0 to 15 do
        if Round(eta_constant[i,j,k].x)<>0 then
         StringGrid1.Cells[j+1,i+1]:=IntToStr(Round(eta_constant[i,j,k].x))+'*e'+IntToStr(k);
     end;
//Output of the sedenion multiplication table into the file.
    IniFile:=TIniFile.Create(GetCurrentDir+'\IniFile.ini');
    array_str:='';
    for i:=0 to 15 do
    begin
      array_str:=array_str+Format('%12s',[chr(36)+'e_{'+IntTostr(i)+'}'+chr(36)]);
      if i<15 then
       array_str:=array_str+'␣&␣';
    end;
```

```
    IniFile.WriteString('Вариант␣','0000000*',array_str+'\\␣\hline');
    for i:=0 to 15 do
    begin
       array_str:='';
       for j:=0 to 15 do
       begin
         for k:=0 to 15 do
           if Round(eta_constant[i,j,k].x)<>0  then
             array_str:=array_str+Format('%12s',[IntToStr(Round(eta_constant[i,j,k].x))+

                        'x'+chr(36)+'e_{'+IntToStr(k)+'}'+chr(36)]);
         if j<15 then
          array_str:=array_str+'␣&␣';
       end;
       if i<10 then
        IniFile.WriteString('Вариант␣','0'+chr(36)+'e_{'+IntTostr(i)+'}'+chr(36),array_str+'\\␣\hline')
       else
        IniFile.WriteString('Вариант␣',chr(36)+'e_{'+IntTostr(i)+'}'+chr(36),array_str+'\\␣\hline');
    end;
 end;
end.
//
```